\documentclass[twocolumn, aip, cha, graphicx,  reprint, numerical, nofootinbib, groupedaddress]{revtex4-1}
\usepackage{amsmath,amssymb,amsfonts,latexsym,parskip,mathtools}
\usepackage{graphicx, subcaption}
\usepackage[justification=justified,singlelinecheck=false]{caption}
\usepackage[figuresright]{rotating} 
\usepackage[hidelinks]{hyperref} 
\usepackage{titlesec}
\usepackage[titletoc,toc,title]{appendix}

\usepackage{sidecap}

\usepackage[usenames,dvipsnames,svgnames]{xcolor}

\hypersetup{
    colorlinks,
    linkcolor={blue!50!black},
    citecolor={blue!50!black},
    urlcolor={blue!50!black}
}

\usepackage{tabularx}
\usepackage{hhline}
\newcolumntype{Y}{>{\centering\arraybackslash}X}

\usepackage{nth}

\usepackage{color}

\newcolumntype{b}{X}
\newcolumntype{s}{>{\hsize=.5\hsize}X}

\usepackage{afterpage}
\makeatletter
\newcommand{\oneraggedpage}{\let\mytextbottom\@textbottom
  \let\mytexttop\@texttop
  \raggedbottom
  \afterpage{%
  \global\let\@textbottom\mytextbottom
  \global\let\@texttop\mytexttop}}
  
\allowdisplaybreaks[1]

\usepackage{comment}

\newcommand{\lb}{\left (}
\newcommand{\rb}{\right )}

\newcommand{\lsq}{\left [}
\newcommand{\rsq}{\right ]}

\newcommand{\RA}{\quad \Rightarrow \quad}

\newcommand{\sechn}[2]{\mathrm{sech}^{#1} \lb #2 \rb}
\newcommand{\tanhn}[2]{\mathrm{tanh}^{#1} \lb #2 \rb}
\newcommand{\dd}[1]{\; \mathrm{d} #1}

\renewcommand{\O}[1]{\mathcal{O} \lb #1 \rb}

\begin{document}

\title{On radiating solitary waves in bi-layers with delamination and coupled Ostrovsky equations}\thanks{Accepted for publication in \textbf{Chaos (2017): http://dx.doi.org/10.1063/1.4973854}}
\author{K. R. Khusnutdinova}
\thanks{
	Corresponding author: K. R. Khusnutdinova. Tel: +44 (0)1509 228202. 
	\textit{E-mail:} K.Khusnutdinova@lboro.ac.uk
}
\affiliation{
	Department of Mathematical Sciences, Loughborough University, Loughborough, LE11 3TU, United Kingdom
}
\author{M. R. Tranter}
\affiliation{
	Department of Mathematical Sciences, Loughborough University, Loughborough, LE11 3TU, United Kingdom
}

\begin{abstract}
We study the scattering of a long longitudinal radiating bulk strain solitary wave in the delaminated area of a two-layered elastic structure with soft (`imperfect') bonding between the layers within the scope of the coupled Boussinesq equations.  The direct numerical modelling of this and similar problems is challenging and has natural limitations. We develop a semi-analytical approach, based on the use of several matched asymptotic multiple-scale expansions and averaging with respect to the fast space variable, leading to the coupled Ostrovsky equations in bonded regions and uncoupled Korteweg-de Vries equations in the delaminated region.
We show that the semi-analytical approach agrees well with direct numerical simulations and  use it to study the nonlinear dynamics and scattering of the radiating solitary wave in a wide range of bi-layers with delamination. The results indicate that radiating solitary waves could help us to control the integrity of layered structures with imperfect interfaces.
\end{abstract}


\maketitle

\begin{quotation}
\bf
Long longitudinal bulk strain solitary waves observed in elastic waveguides, such as rods, bars,  plates and shells, can be modelled with Boussinesq-type equations. 
Radiating solitary waves, that is solitary waves radiating a co-propagating one-sided oscillatory tail, 
emerge in layered elastic waveguides with soft (`imperfect') bonding between the layers. 
In this paper we study the scattering of a radiating solitary wave in delaminated areas of imperfectly bonded bi-layers within the scope of the coupled Boussinesq equations. We develop direct and semi-analytical numerical approaches
and demonstrate that radiating solitary waves undergo changes which could be used to control the quality of the interfaces.

\end{quotation}

\section{Introduction}
The discovery of solitons as extremely stable localised coherent structures \cite{ZK} is intrinsically linked with the discovery of the Inverse Scattering Transform (IST) for the Korteweg-de Vries (KdV) equation - the method for the solution of a large class of initial-value problems on the infinite line. \cite{GGKM}
The latter has shown that solitons, when present, constitute the main part of the long-time asymptotics of initial-value problems for localised initial data, and this is the reason why solitons proved to be a very important part of the physical world we live in, across all scales. \cite{SS, NM, Hoefer, Dreiden15}

Initially developed as a purely analytical technique, \cite{AS, ZS} in recent years the IST formed the basis for the development of efficient numerical approaches to the analysis of nonlinear problems, most notably within the framework of another famous integrable model, the Nonlinear Schr\"odinger (NLS) equation. \cite{ZS, Osborne, SCDA} An efficient IST-based numerical approach to solving the KdV equation was also developed. \cite{TOD}

Recently, the method has found  a new application  in our studies of the scattering of long longitudinal bulk strain solitons in a symmetric perfectly bonded layered bar with delamination. \cite{KS, KT} This condensed matter problem can be viewed as an analogue to the fluid mechanics problem of calculating the reflected and transmitted waves when a surface or internal soliton passes through an area of rapid depth variation. \cite{Pelinovsky, Tappert, Johnson, Djordjevic, Grimshaw08} The theoretical predictions \cite{KS} agreed well with experimental studies, \cite{Dreiden10} and were also confirmed by numerical simulations. \cite{KT}  Long longitudinal bulk strain solitary waves were experimentally observed in various elastic waveguides, including rods, bars, plates and shells, and modelled using Boussinesq-type equations. \cite{Samsonov_book, Porubov_book, Dreiden10, Dreiden15}
The exceptional stability of bulk strain solitons \cite{Samsonov_book, SDS_2008, SSB_2016} makes them an attractive candidate for the introscopy of  layered structures, in addition to the existing methods. \cite{Dreiden12, SBDPS}

The dynamical behaviour of layered structures depends not only on the properties of the bulk material, but also on the type of the bonding between the layers. In particular, if the materials of the layers have similar properties and the bonding between the layers is sufficiently soft (`imperfect bonding'), then the bulk strain soliton is replaced with a {\it radiating solitary wave}, a solitary wave with a co-propagating oscillatory tail. \cite{KSZ} The radiating strain solitary wave has recently been observed in laboratory experiments. \cite{Dreiden12} More generally, experimental studies of the excitation of the resonant radiation by localised waves have been a prominent theme in nonlinear optics and a number of other physical settings, see, for example, the reviews \cite{SG, CTMK} and the references therein.

Indeed it was shown, within the framework of a complex lattice model,  that long nonlinear longitudinal bulk strain waves in a bi-layer with a sufficiently soft bonding can be modelled with a system of coupled regularised Boussinesq (cRB) equations \cite{KSZ} (in non-dimensional and scaled form):
\begin{align}
f_{tt} - f_{xx} &= \frac 12 (f^2)_{xx} + f_{ttxx} - \delta (f-g)\,, \nonumber \\
g_{tt} - c^2 g_{xx} &= \frac 12  \alpha  (g^2)_{xx} + \beta  g_{ttxx} + \gamma  (f-g)\,. 
\label{fg}
\end{align}
Here, $f$ and $g$ denote the  longitudinal strains in the layers, while the coefficients $c, \alpha, \beta, \delta, \gamma$ are defined by the physical and geometrical parameters of the problem \cite{KSZ} (see Section II for details). 

\begin{figure}[ht]
	\begin{subfigure}[t]{0.40\textwidth}
		\centering
		\includegraphics[width=\textwidth, trim = 10mm 0mm 10mm 0mm]{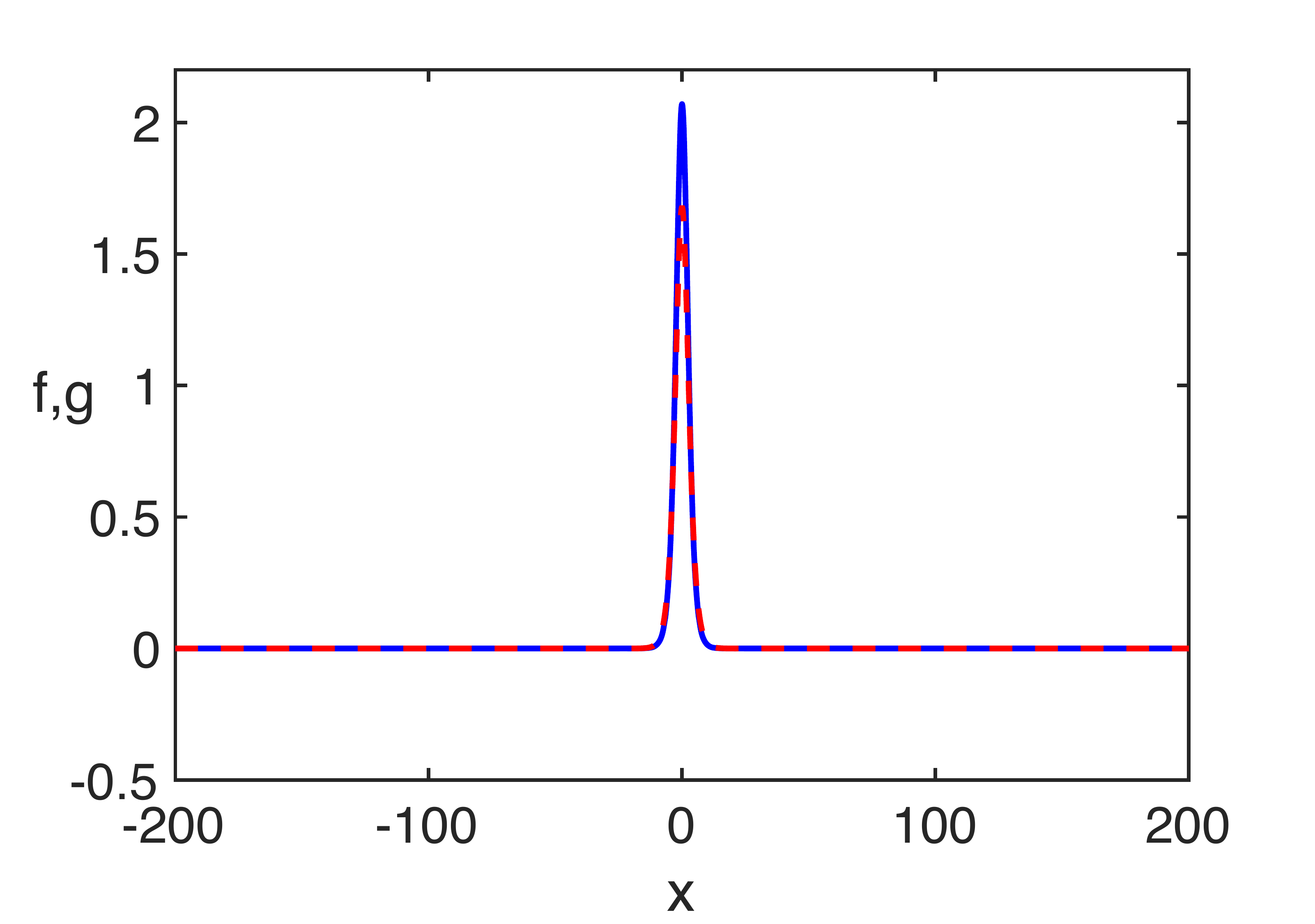}
		\caption{ }
	\end{subfigure}
	\\
	\begin{subfigure}[t]{0.40\textwidth}
		\centering
		\includegraphics[width=\textwidth, trim = 10mm 0mm 10mm 0mm]{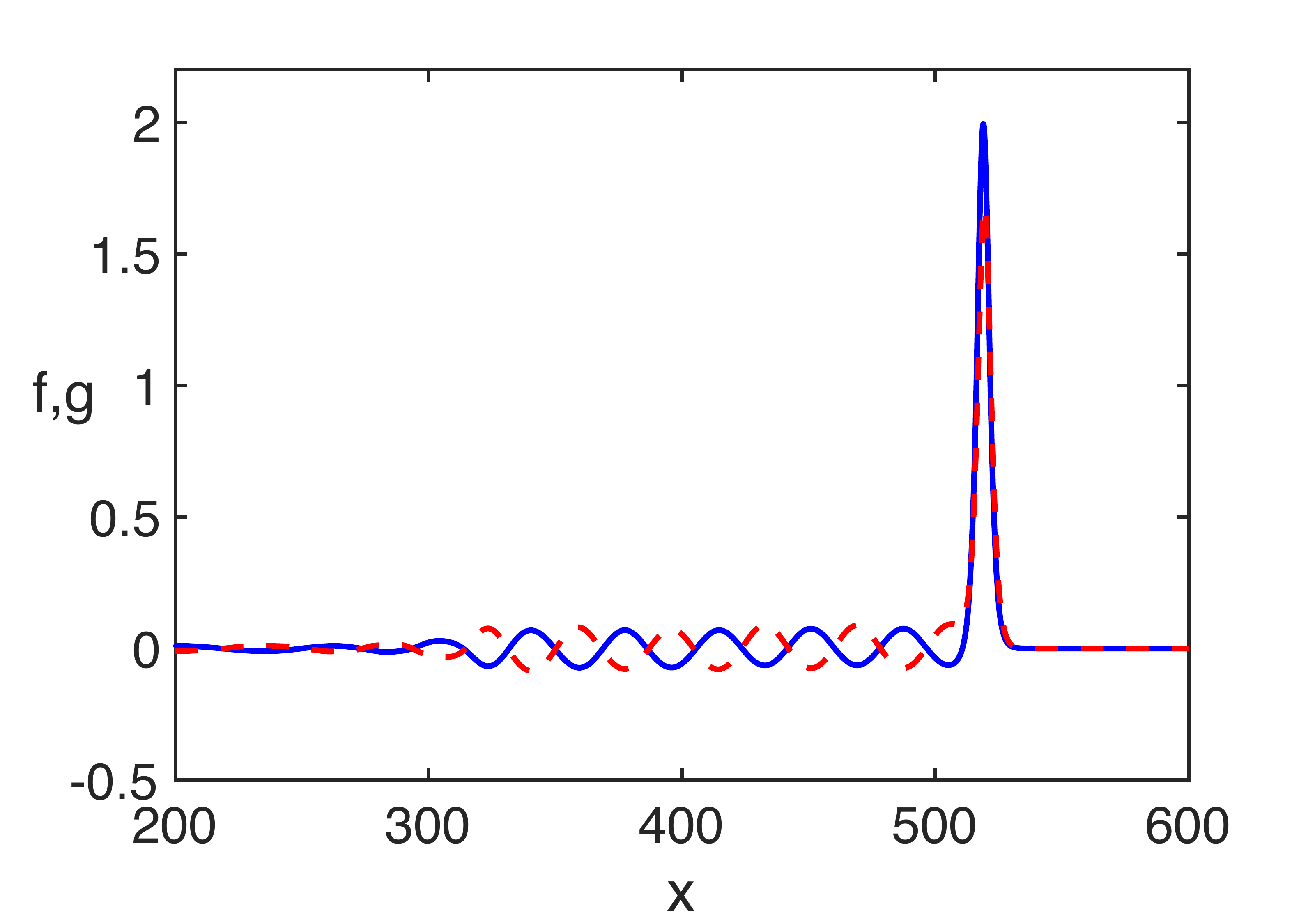}
		\caption{ }
	\end{subfigure}
	\caption{\small Typical generation of radiating solitary waves in 
	system \eqref{fg}, from pure solitary wave initial conditions, for $f$ (solid line) and $g$ (dashed line). Here $c=1.05$, $\alpha = \beta = 1.05$.  (a) Initial condition at $t = 0$:  pure solitary wave solution with $\delta = \gamma = 0$ and $v=1.3$.  (b) Radiating solitary wave solution with  $\delta = \gamma = 0.01$  at $t = 400$.}
	\label{pure_solitary_wave}
\end{figure}

In the symmetric case ($c=\alpha=\beta=1$) system (\ref{fg}) admits the reduction $g=f$, where $f$ satisfies the equation
\begin{equation}
f_{tt} - f_{xx} = \frac 12 (f^2)_{xx} + f_{ttxx}\,.
\label{f_intro}
\end{equation}
The Boussinesq equation (\ref{f_intro}) has particular solitary wave solutions:
{\small
${\displaystyle f = 3 (v^2 - 1) \ \sechn{2}{\frac{x - v t}{\Lambda}}\hspace{-0.28em}, 
\quad  \Lambda = \frac{2 v}{\sqrt{v^2 - 1}},}$
}where $v$ is the speed of the wave. However, in the system of cRB equations (\ref{fg}), when the characteristic speeds of the linear waves in the layers are close (i.e. $c$ is close to 1), these pure solitary wave solutions are replaced with radiating solitary waves, \cite{KSZ, KM} that is solitary waves radiating a co-propagating one-sided oscillatory tail. \cite{BGK, Shrira, Bona} Figure \ref{pure_solitary_wave} illustrates the pure solitary wave solutions of system (\ref{fg}) with $\delta = \gamma = 0$ (for a fixed value of $v$), and subsequent evolution of this initial condition into a radiating solitary wave.
Radiating solitary waves have been extensively studied in the context of perturbed KdV equations, coupled KdV systems, perturbed NLS equations, and coupled NLS systems. \cite{Boyd,Grimshaw,GI,GJ,Karpman,Lombardi,VB} The radiating solitary waves emerge due to a resonance between a solitary wave and some linear wave, which can be deduced from the analysis of the relevant linear dispersion relation.

\begin{figure}[ht]
\begin{center}
\includegraphics[width=0.45 \textwidth]{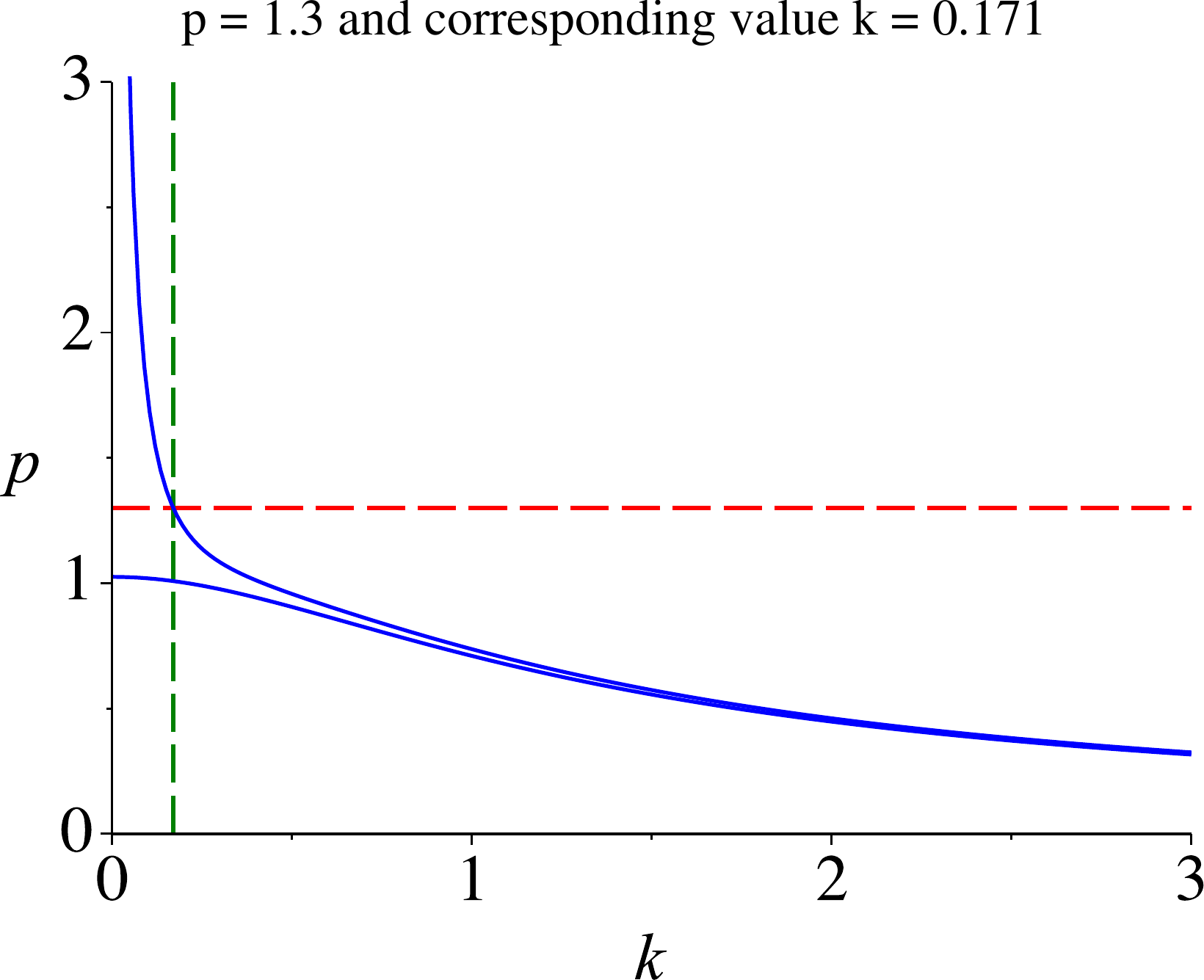}
\end{center}
\caption{\small Two branches of the linear dispersion curve of 
system (\ref{fg}) for $c=1.05$, $\alpha = \beta = 1.05$,  $\delta = \gamma = 0.01$ and 
a 
resonance for $p= 1.3$ (horizontal line).}
\label{disp_curves_intro}
\end{figure}

When considering the linear dispersion relation for  the system (\ref{fg}), it is assumed that the coefficients in (\ref{fg}) are perturbed compared to the symmetric case, but remain positive. \cite{KSZ} The dispersion relation has the form
\begin{equation}
[k^2 (1-p^2) - k^4 p^2 + \delta] [k^2 (c^2-p^2) - \beta k^4 p^2 + \gamma] = \gamma \delta,
\label{disp}
\end{equation}
where $k$ is the wavenumber and $p$ is the phase speed. A typical plot of (\ref{disp})
is  shown in Figure \ref{disp_curves_intro}. A significant difference with the linear dispersion curve of the reduction (\ref{f_intro})  is the appearance of the second branch, remaining above the first branch for all $k$, and going to infinity as $k \to 0$, while approaching zero as $k \to \infty$. 
The pure solitary waves  of the single Boussinesq equation (\ref{f_intro})   arise as a bifurcation from wavenumber $k = 0$ of the linear wave spectrum, when there is no possible resonance between the speed $v$ of the solitary wave and the phase speed $p$ of some linear wave. Radiating solitary waves  arise  in the case when there is a possible resonance for some finite non-zero value of $k$. For example, a possible resonance is shown in Figure \ref{disp_curves_intro} for $v=p=1.3$.

The aim of this paper is to study the scattering of this type of solitary wave in delaminated areas of imperfectly bonded layered structures (see Figure 3). We develop a semi-analytical approach which leads to coupled Ostrovsky equations in bonded regions of a bi-layer, and to uncoupled Korteweg-de Vries equations in the delaminated area. The Ostrovsky equation was originally derived to describe long surface and internal waves in a rotating ocean, \cite{Ostrovsky, review} but recently it transpired that the equation, as well as the coupled Ostrovsky equations, can also describe nonlinear strain waves in layered elastic waveguides with soft interfaces. \cite{KM}
We also develop a direct numerical approach and verify that the semi-analytical method produces the correct results in the cases where we can use both methods. However the direct numerical simulations are expensive, therefore we then use our semi-analytical method to study the scattering of radiating solitary waves in a wide range of complex imperfectly bonded bi-layers with delamination, giving an elaborate description of the possible dynamical effects.

The paper is organised as follows. In Section \ref{sec:Form} we state the problem formulation for the generation and the scattering of a radiating solitary wave in an imperfectly bonded bi-layer with delamination. In Section \ref{sec:WNL} we develop a weakly nonlinear solution of this scattering problem and discuss the related semi-analytical approach. In Section \ref{sec:InfDelam}, in a case study, we compare the results obtained using the semi-analytical approach with the results of direct numerical simulations, and show that the results are in good agreement. We then continue to use the semi-analytical approach to study the scattering of radiating solitary waves for a wide range of configurations of the complex structure. We summarise our findings and discuss the results in Section \ref{sec:Conc}.

\section{Problem Formulation}
\label{sec:Form}
\begin{figure}[!ht]
\center
\includegraphics[width=0.45\textwidth, trim=0mm 0mm 0mm 10mm]{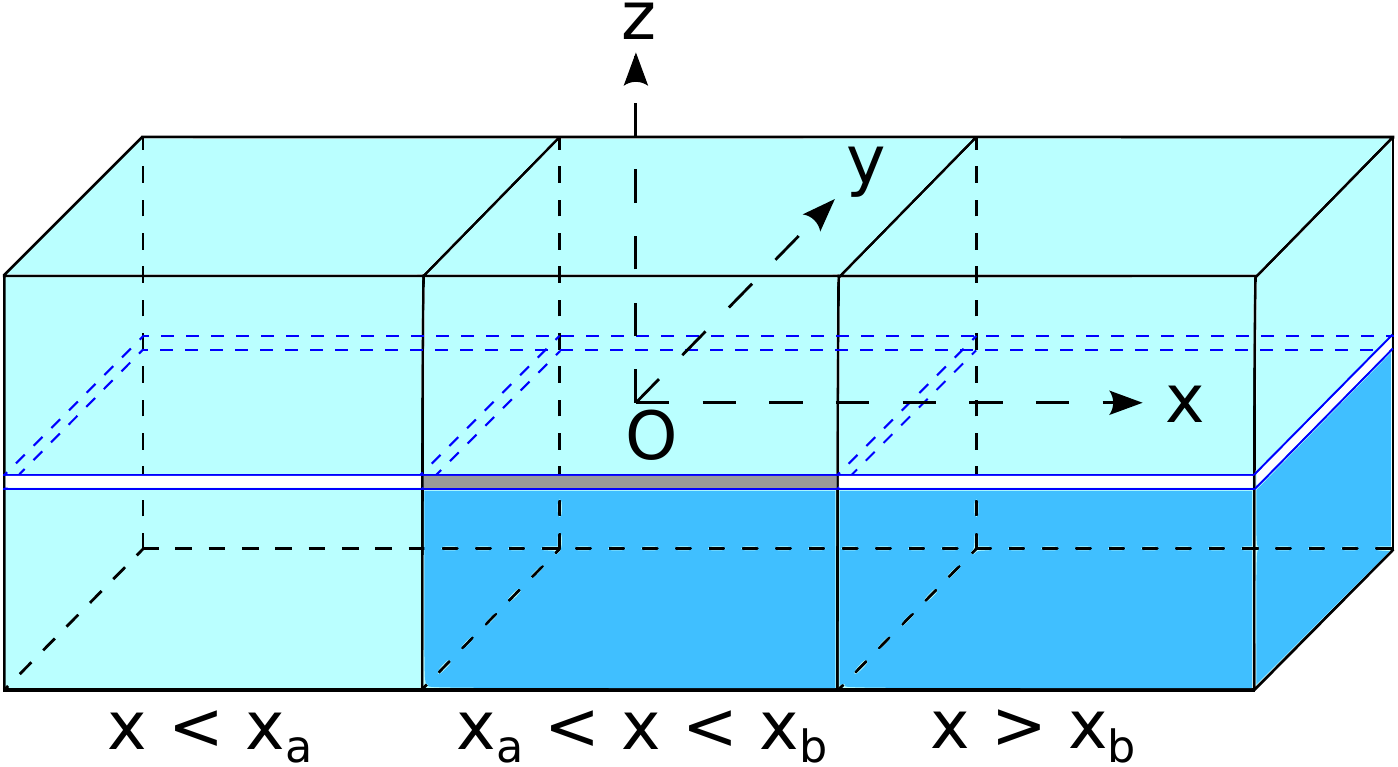}
\caption{Bi-layer with two homogeneous layers for $x < x_a$, a bonded two-layered section for $x_a < x < x_b$ and a delaminated section for $x > x_b$.}
\label{fig:DelamBar}
\vspace{-1em}
\end{figure}

We consider the generation and the scattering of a long radiating solitary wave in a two-layered imperfectly bonded bi-layer with delamination, shown in Figure \ref{fig:DelamBar}.
The model is inspired by the  experimental setup. \cite{Dreiden12} Two identical homogeneous layers (the section on the left) are `glued'  to a two-layered structure with soft bonding between its layers (in the middle), followed with a delaminated section (on the right). The materials in the bi-layer are assumed to have close properties, leading to the generation of a radiating solitary wave in the bonded section. \cite{KSZ} We study the scattering of this wave by the subsequent delaminated region.

The mathematical problem formulation consists of the following sets of scaled regularised non-dimensional equations in the respective sections of the complex waveguide: \cite{KS, KSZ, KT}
\begin{align}
u_{tt}^{(1)} - u_{xx}^{(1)} &= \epsilon \lsq -12 u_{x}^{(1)} u_{xx}^{(1)} + 2 u_{ttxx}^{(1)} \rsq, \notag \\
w_{tt}^{(1)} - w_{xx}^{(1)} &= \epsilon \lsq -12 w_{x}^{(1)} w_{xx}^{(1)} + 2 w_{ttxx}^{(1)} \rsq \label{syshomog}
\end{align} 
for $x < x_a$; 
\begin{align}
u_{tt}^{(2)} - u_{xx}^{(2)} &= \epsilon \lsq -12 u_{x}^{(2)} u_{xx}^{(2)} + 2 u_{ttxx}^{(2)} \right. \notag \\
&\left. \qquad -~ \delta \lb u^{(2)} - w^{(2)} \rb \rsq, \notag \\
w_{tt}^{(2)} - c^2 w_{xx}^{(2)} &= \epsilon \lsq -12 \alpha w_{x}^{(2)} w_{xx}^{(2)} + 2 \beta w_{ttxx}^{(2)} \right. \notag \\
&\left. \qquad +~ \gamma \lb u^{(2)} - w^{(2)} \rb \rsq \label{syscoup} 
\end{align}
for $x_a < x < x_b$; and
\begin{align}
u_{tt}^{(3)} - u_{xx}^{(3)} &= \epsilon \lsq -12 u_{x}^{(3)} u_{xx}^{(3)} + 2 u_{ttxx}^{(3)} \rsq, \notag \\
w_{tt}^{(3)} - c^2 w_{xx}^{(3)} &= \epsilon \lsq -12 \alpha w_{x}^{(3)} w_{xx}^{(3)} + 2 \beta w_{ttxx}^{(3)} \rsq \label{sysdelam}
\end{align}
for $x > x_b$. Here, the functions $u^{(i)}(x,t)$ and $w^{(i)}(x,t)$ describe longitudinal displacements in the `top' and `bottom' layers of the three sections of the waveguide, respectively. 
The values of the constants $\alpha$, $\beta$ and $c$ depend on the physical and geometrical properties of the waveguide, while the constants $\delta$ and $\gamma$ depend on the properties of the soft bonding layer, and $\epsilon$ is a small amplitude parameter. \cite{KSZ}

These equations are complemented with continuity conditions at the interfaces between the sections: continuity of longitudinal displacement
\begin{align}
u^{(1)} |_{x=x_a} &= u^{(2)} |_{x=x_a}, & w^{(1)} |_{x=x_a} &= w^{(2)} |_{x=x_a}; \label{cont_a} \\
u^{(2)} |_{x=x_b} &= u^{(3)} |_{x=x_b}, &w^{(2)} |_{x=x_b} &= w^{(3)} |_{x=x_b}; \label{cont_b}
\end{align}
and continuity of normal stress
{\small
\begin{align*}
&\left. u_{x}^{(1)} + \epsilon \lsq -6 \lb u_{x}^{(1)} \rb^2 + 2 u_{ttx}^{(1)} \rsq \right |_{x=x_a} = \\
&\left. u_{x}^{(2)} + \epsilon \lsq -6 \lb u_{x}^{(2)} \rb^2 + 2 u_{ttx}^{(2)} \rsq \right |_{x=x_a},
\end{align*}
\begin{align}
&\left. w_{x}^{(1)} + \epsilon \lsq -6 \lb w_{x}^{(1)} \rb^2 + 2 w_{ttx}^{(1)} \rsq \right |_{x=x_a} = \notag \\
&\left. c^2 w_{x}^{(2)} + \epsilon \lsq -6 \alpha \lb w_{x}^{(2)} \rb^2 + 2 \beta w_{ttx}^{(2)} \rsq \right |_{x=x_a}; \label{cont2_a}
\end{align}
and
\begin{align*}
&\left. u_{x}^{(2)} + \epsilon \lsq -6 \lb u_{x}^{(2)} \rb^2 + 2 u_{ttx}^{(2)} \rsq \right |_{x=x_b} =  \\
&\left. u_{x}^{(3)} + \epsilon \lsq -6 \lb u_{x}^{(3)} \rb^2 + 2 u_{ttx}^{(3)} \rsq \right |_{x=x_b},
\end{align*}
\begin{align}
&\left. c^2 w_{x}^{(2)} + \epsilon \lsq -6 \alpha \lb w_{x}^{(2)} \rb^2 + 2 \beta w_{ttx}^{(2)} \rsq \right |_{x=x_b} = \notag \\
&\left. c^2 w_{x}^{(3)} + \epsilon \lsq -6 \alpha \lb w_{x}^{(3)} \rb^2 + 2 \beta w_{ttx}^{(3)} \rsq \right |_{x=x_b}; \label{cont2_b}
\end{align}
}as well as some natural initial and boundary conditions, which will be imposed later.

\section{Weakly Nonlinear Solution}
\label{sec:WNL}

Differentiating the governing equations with respect to $x$ and denoting $f^{(i)} = u_x^{(i)}$ and $g^{(i)} = w_x^{(i)}$, we obtain the equations `in strains':
\begin{align}
f_{tt}^{(1)} - f_{xx}^{(1)} &= \epsilon \lsq -6 \lb f^{(1)} \rb^2+ 2 f_{tt}^{(1)} \rsq _{xx}, \notag \\
g_{tt}^{(1)} - g_{xx}^{(1)} &= \epsilon \lsq -6 \lb g^{(1)} \rb^2 + 2 g_{tt}^{(1)} \rsq _{xx} \label{syshomogdiff} 
\end{align}
for $x < x_a$;
\begin{align}
f_{tt}^{(2)} - f_{xx}^{(2)} &= \epsilon \lsq -6 \lb f^{(2)} \rb^2 + 2  f_{tt}^{(2)} \rsq _{xx} \notag \\
& \qquad -~ \epsilon \delta \lb f^{(2)} - g^{(2)} \rb , \notag \\
g_{tt}^{(2)} - c^2 g_{xx}^{(2)} &= \epsilon \lsq -6 \alpha \lb g^{(2)} \rb^2 + 2 \beta g_{tt}^{(2)} \rsq_{xx} \notag \\
& \qquad +~ \epsilon \gamma \lb f^{(2)} - g^{(2)} \rb  \label{syscoupdiff} 
\end{align}
for $x_a < x < x_b$; and 
\begin{align}
f_{tt}^{(3)} - f_{xx}^{(3)} &= \epsilon \lsq -6 \lb f^{(3)} \rb^2 + 2 f_{tt}^{(3)} \rsq _{xx}, \notag \\
g_{tt}^{(3)} - c^2 g_{xx}^{(3)} &= \epsilon \lsq -6 \alpha \lb g^{(3)} \rb^2 + 2 \beta g_{tt}^{(3)} \rsq  _{xx} \label{sysdelamdiff}
\end{align}
for $x > x_b$. 
 
To find the weakly nonlinear solution of the complicated scattering problem we consider the equations \eqref{syshomogdiff} - \eqref{sysdelamdiff}. We use several asymptotic multiple-scale expansions, and develop a space-averaging method instead of the time-averaging method used for the homogeneous initial-value problem. \cite{KM} All functions present in our expansions and their derivatives are assumed to be bounded and sufficiently rapidly decaying at infinity (these assumptions agree with our numerical simulations). 

In the regions where the behaviour is governed by uncoupled regularised Boussinesq equations, the previous work  shows that to leading order the weakly nonlinear solution  is described by the appropriate KdV equations. \cite{KS, KT} Therefore, we will omit the majority of the derivation in these regions and instead focus on the coupled regularised Boussinesq equations in \eqref{syscoupdiff}. Finally we will use the continuity conditions (\ref{cont_a}) - (\ref{cont2_b}) to obtain `initial conditions' for the derived leading order equations.

\subsection{First Region: two homogeneous layers}
In the first region, the equation is identical in both homogeneous layers and therefore we assume the same incident wave in both, and consider asymptotic multiple-scale expansions of the type
\begin{align*}
f^{(1)} = I \lb \xi, X \rb + R^{(1)} \lb \eta, X \rb + \epsilon P^{(1)} \lb \xi, \eta, X \rb + \O{\epsilon^2}, \\
g^{(1)} = I \lb \xi, X \rb + G^{(1)} \lb \eta, X \rb + \epsilon Q^{(1)} \lb \xi, \eta, X \rb + \O{\epsilon^2}, 
\end{align*}
where the characteristic variables are given by $\xi = x - t$, $\eta = x + t$,  and  the slow variable $X = \epsilon x$. Here, the functions $I$ and $R^{(1)}$, $G^{(1)}$ represent the leading order incident and reflected waves respectively and $P^{(1)}, Q^{(1)}$ are the higher order corrections. Substituting the asymptotic expansion into the first equation in \eqref{syshomogdiff}, the system is satisfied at leading order, while at $\O{\epsilon}$ we have
\begin{align}
&-2 P^{(1)}_{\xi \eta} = \lb I_{X} - 6 I I_{\xi} + I_{\xi \xi \xi} \rb_{\xi} + \lb R^{(1)}_{X} - 6 R^{(1)} R^{(1)}_{\eta} \right . \nonumber \\
&+ \left .  R^{(1)}_{\eta \eta \eta} \rb_{\eta}  - 6 \lb 2 I_{\xi} R^{(1)}_{\eta} + R^{(1)}_{\eta \eta} I + I_{\xi \xi} R^{(1)} \rb,
\label{P1exp}
\end{align}
and a similar equation can be obtained for the second layer.
We average equation \eqref{P1exp} with respect to the fast space variable $x$ using  $\displaystyle \lim_{\chi \rightarrow - \infty} \frac{1}{x_a - \chi} \int_{\chi}^{x_a} \dots \; \dd{x},$ in the reference frame moving with the linear speed of right- and left-propagating waves (at constant $\xi$ or $\eta$).  Assuming that all functions and their derivatives remain bounded (in order to avoid secular terms in asymptotic expansions) and decay sufficiently rapidly at infinity we have, for example at constant $\xi$,
{\small 
\begin{eqnarray}
\lim_{\chi \rightarrow - \infty} \frac{1}{x_a - \chi} \int\displaylimits_{\chi}^{x_a} P^{(1)}_{\xi \eta} \dd{x} = \lim_{\chi \rightarrow -\infty} \frac{1}{2 (x_a - \chi ) } \int\displaylimits_{2 \chi - \xi}^{2 x_a - \xi}  P^{(1)}_{\xi \eta} \dd{\eta} \nonumber \\
=  \lim_{\chi \rightarrow -\infty}  \frac{1}{2 (x_a - \chi ) } \lsq P^{(1)}_{\xi} \rsq_{2 \chi - \xi}^{2 x_a - \xi} = 0. \qquad \label{aveximinus}
\end{eqnarray}
}The same result can be obtained for $P^{(1)}$ at constant $\eta$. 
Therefore, we can average \eqref{P1exp} at constant $\xi$ to obtain
\begin{equation}
\lb I_{X} - 6 I I_{\xi} + I_{\xi \xi \xi} \rb_{\xi} = 0. 
\label{Ieqint}
\end{equation}
Similarly, averaging \eqref{P1exp} with respect to $x$ at constant $\eta$ gives
\begin{equation}
\lb R^{(1)}_{X} - 6 R^{(1)} R^{(1)}_{\eta} + R^{(1)}_{\eta \eta \eta} \rb_{\eta} = 0.
\end{equation}
In each case, we can integrate with respect to the relevant characteristic variable and, recalling that there are no waves at infinity, we obtain
\begin{align}
I_{X} - 6 I I_{\xi} + I_{\xi \xi \xi} &= 0, \label{Ieq} \\
R^{(1)}_{X} - 6 R^{(1)} R^{(1)}_{\eta} + R^{(1)}_{\eta \eta \eta} &= 0. \label{Req}
\end{align}
Substituting \eqref{Ieq} and \eqref{Req} into \eqref{P1exp} and integrating with respect to the characteristic variables, we obtain an expression for $P^{(1)}$ of the form
\begin{align}
P^{(1)} &= 3 \lb 2 I R^{(1)} + R^{(1)}_{\eta} \int I \dd{\xi} + I_{\xi} \int R^{(1)} \dd{\eta} \rb \notag \\
&\quad+ \phi_1^{(1)} \lb \xi, X \rb + \psi_1^{(1)} \lb \eta, X \rb,
\label{P1eq}
\end{align}
where $\phi_1^{(1)}$, $\psi_1^{(1)}$ are arbitrary functions. 

Similarly, we obtain the equations 
\begin{align}
G^{(1)}_{X} &- 6 G^{(1)} G^{(1)}_{\eta} + G^{(1)}_{\eta \eta \eta} = 0, \label{Geq} \\
Q^{(1)} &= 3 \lb 2 I G^{(1)} + G^{(1)}_{\eta} \int I \dd{\xi} + I_{\xi} \int G^{(1)} \dd{\eta} \rb \notag \\
&\quad + \phi_2^{(1)} \lb \xi, X \rb + \psi_2^{(1)} \lb \eta, X \rb,
\label{Q1eq}
\end{align}
for the waves in the second layer, in addition to \eqref{Ieq}.

The first radiation condition requires that the solution to the problem should not change the incident wave. \cite{KS, KT} For the case of an incident solitary wave this implies that $\phi_1^{(1)} = 0$ and $\phi_2^{(1)} = 0$. 

\subsection{Second region: bi-layer with soft bonding}
We assume that the layers have close properties, so that $c-1 = \O{\epsilon}$. In this case the cRB equations admit solutions in the form of radiating solitary waves. \cite{KSZ, KM}
Thus, we assume that
\begin{equation*}
c - 1 = \O{\epsilon} \RA \frac{c^2 - 1}{\epsilon} = \O{1},
\end{equation*}
and therefore we can make the rearrangement
\begin{align}
g_{tt}^{(2)} - g_{xx}^{(2)} =&~ \epsilon \lsq -6 \alpha  \lb g^{(2)} \rb^2  + 2 \beta g_{tt}^{(2)} + \frac{c^2 - 1}{\epsilon} g^{(2)} \rsq _{xx}  \notag \\
&  + \epsilon \gamma \lb f^{(2)} - g^{(2)} \rb.
\label{gmod}
\end{align}
This allows us to use one set of characteristic variables for $f^{(2)}$ and $g^{(2)}$. Let us assume that there is a weakly nonlinear solution to \eqref{syscoupdiff} of the form
\begin{align*}
f^{(2)} &= T^{(2)} \lb \xi, X \rb + R^{(2)} \lb \eta, X \rb + \epsilon P^{(2)} \lb \xi, \eta, X \rb + \O{\epsilon^2},  \\
g^{(2)} &= S^{(2)} \lb \xi, X \rb + G^{(2)} \lb \eta, X \rb + \epsilon Q^{(2)} \lb \xi, \eta, X \rb + \O{\epsilon^2}.
\end{align*}
The characteristic variables $\xi$, $\eta$ and $X$ are the same as before, $T^{(2)}$ and $S^{(2)}$ represent the transmitted waves in the second section of the bar, where $T$ is for the top layer and $S$ is for the bottom layer. Similarly, $R^{(2)}$ and $G^{(2)}$ are the reflected waves, and the higher order corrections in this section are given by $P^{(2)}$ and $Q^{(2)}$, for the top and bottom layers respectively.

The solution is considered over the period of time until the waves reflected from the boundary $x = x_b$, between the second and the third region, reach the boundary $x = x_a$, between the first and the second region. Moreover, the boundary $x = x_b$ is assumed to be sufficiently far away from the boundary $x = x_a$, allowing us to use the averaging $\displaystyle \lim_{x_b \to \infty} \frac{1}{x_b - x_a} \int_{x_a}^{x_b} \dots \ dx.$ Substituting the asymptotic expansion into the equation for $f^{(2)}$ in \eqref{syscoupdiff} the equation is satisfied at leading order, while at $\O{\epsilon}$ we have
{\small \begin{align}
&-2P^{(2)}_{\xi \eta} = \lb T^{(2)}_{X} - 6 T^{(2)} T^{(2)}_{\xi} + T^{(2)}_{\xi \xi \xi} \rb_{\xi} + \lb R^{(2)}_{X} - 6 R^{(2)} R^{(2)}_{\eta} \right . \nonumber\\
&+ \left . R^{(2)}_{\eta \eta \eta} \rb_{\eta} - 6 \lb 2 T^{(2)}_{\xi} R^{(2)}_{\eta} + T^{(2)} R^{(2)}_{\eta \eta} + T^{(2)}_{\xi \xi} R^{(2)} \rb   \nonumber \\
&- \frac{\delta}{2} \lb T^{(2)} - S^{(2)} + R^{(2)} - G^{(2)} \rb.
\label{P2exp}
\end{align}
}For the equation governing $g^{(2)}$ we have
{\small \begin{align}
&-2Q^{(2)}_{\xi \eta} = \lb S^{(2)}_{X} + \frac{c^2 - 1}{2 \epsilon} S^{(2)}_{\xi} - 6 \alpha S^{(2)} S^{(2)}_{\xi} + \beta S^{(2)}_{\xi \xi \xi} \rb_{\xi} \nonumber \\
 &+ \lb G^{(2)}_{X} + \frac{c^2 - 1}{2 \epsilon} G^{(2)}_{\eta}- 6 \alpha G^{(2)} G^{(2)}_{\eta} + \beta G^{(2)}_{\eta \eta \eta} \rb_{\eta} \nonumber \\
   &- 6 \alpha \lb 2 S^{(2)}_{\xi} G^{(2)}_{\eta}  + S^{(2)} G^{(2)}_{\eta \eta} + S^{(2)}_{\xi \xi} G^{(2)} \rb \nonumber \\
&+ \frac{\gamma}{2} \lb T^{(2)} - S^{(2)} + R^{(2)} - G^{(2)} \rb.
\label{Q2exp}
\end{align}
}We average equations \eqref{P2exp} and \eqref{Q2exp} with respect to the fast space variable $x$ as defined earlier. In each case, we average at constant $\xi$ or $\eta$ and note that $P^{(2)}$ and $Q^{(2)}$ are both zero when averaged at either constant $\xi$ or constant $\eta$. Averaging \eqref{P2exp} and \eqref{Q2exp}  at constant $\xi$ gives
\begin{align}
& \lb T^{(2)}_{X} - 6 T^{(2)} T^{(2)}_{\xi} + T^{(2)}_{\xi \xi \xi} \rb_{\xi} = \frac{\delta}{2} \lb T^{(2)} - S^{(2)} \rb,  \label{T2eq} \\
& \lb S^{(2)}_{X} + \frac{c^2 - 1}{2 \epsilon} S^{(2)}_{\xi} - 6 \alpha S^{(2)} S^{(2)}_{\xi} + \beta S^{(2)}_{\xi \xi \xi} \rb_{\xi} \nonumber \\
& \hspace{3.8cm}   = \frac{\gamma}{2} \lb S^{(2)} - T^{(2)} \rb,  
\label{S2eq}
\end{align}
and therefore \eqref{T2eq} and \eqref{S2eq} form a system of coupled Ostrovsky equations. \cite{KM} We note that the Ostrovsky equation was initially derived to describe long surface and internal waves in a rotating ocean. \cite{Ostrovsky, review} Coupled Ostrovsky equations appear naturally in the description of nonlinear waves in layered waveguides, both solid and fluid. \cite{KM, AGK14}

Similarly, averaging \eqref{P2exp} and \eqref{Q2exp} at constant $\eta$ gives
\begin{align}
&\lb R^{(2)}_{X} - 6 R^{(2)} R^{(2)}_{\eta} + R^{(2)}_{\eta \eta \eta} \rb_{\eta} =  \frac{\delta}{2} \lb R^{(2)} - G^{(2)} \rb, \label{R2eq} \\
& \lb G^{(2)}_{X} + \frac{c^2 - 1}{2 \epsilon} G^{(2)}_{\eta} - 6 \alpha G^{(2)} G^{(2)}_{\eta} + \beta G^{(2)}_{\eta \eta \eta} \rb_{\eta} \nonumber \\
&\hspace{4.4cm} = \frac{\gamma}{2} \lb G^{(2)} - R^{(2)} \rb ,
\label{G2eq}
\end{align}
respectively. Therefore, to leading order, the transmitted and reflected waves are described by two systems of coupled Ostrovsky equations. This result is consistent with the time-averaged derivation. \cite{KM}

Substituting \eqref{T2eq} and \eqref{R2eq} into \eqref{P2exp} and integrating with respect to the characteristic variables, we obtain
\begin{align}
P^{(2)} &= 3 \lb 2 T^{(2)} R^{(2)} + R^{(2)}_{\eta} \int T^{(2)} \dd{\xi} + T^{(2)}_{\xi} \int R^{(2)} \dd{\eta} \rb \notag \\
&\quad + \phi_1^{(2)} \lb \xi, X \rb + \psi_1^{(2)} \lb \eta, X \rb,
\label{P2eq}
\end{align}
where $\phi_1^{(2)}$, $\psi_1^{(2)}$ are arbitrary functions. 

Similarly, substituting \eqref{S2eq} and \eqref{G2eq} into \eqref{Q2exp} and integrating with respect to the characteristic variables, we obtain
\begin{align}
Q^{(2)} &= 3 \alpha \lb 2 S^{(2)} G^{(2)} + G^{(2)}_{\eta} \int S^{(2)} \dd{\xi} + S^{(2)}_{\xi} \int G^{(2)} \dd{\eta} \rb \notag \\
&\quad + \phi_2^{(2)} \lb \xi, X \rb + \psi_2^{(2)} \lb \eta, X \rb,
\label{Q2eq}
\end{align}
where $\phi_2^{(2)}$, $\psi_2^{(2)}$ are arbitrary functions. 

\subsection{Third region: delamination}
\label{sec:R3}
We now consider the third region, where the same bi-layered waveguide does not have a bonding layer, modelling delamination. The motion in this region is governed by two uncoupled regularised Boussinesq equations, but with differing coefficients in each layer. We look for a weakly nonlinear solution to \eqref{sysdelamdiff} of the form
\begin{align*}
f^{(3)} &= T^{(3)} \lb \xi, X \rb + \epsilon P^{(3)} \lb \xi, \eta, X \rb + \O{\epsilon^2},  \\
g^{(3)} &= S^{(3)} \lb \nu, X \rb + \epsilon Q^{(3)} \lb \nu, \zeta, X \rb + \O{\epsilon^2},
\end{align*}
where we now use two sets of characteristic variables $\xi = x - t$, $\eta = x + t$, and $\nu = x - c t$,  $\zeta = x + c t$,  while $X = \epsilon x$. Substituting this into system \eqref{sysdelamdiff} the equation is satisfied at leading order, while at $\O{\epsilon}$ we have
\begin{align}
-2P^{(3)}_{\xi \eta} &= \lb T^{(3)}_{X} - 6 T^{(3)} T^{(3)}_{\xi} + T^{(3)}_{\xi \xi \xi} \rb_{\xi}, \label{P3exp} \\
-2 Q^{(3)}_{\nu \zeta} &= \lb S^{(3)}_{X} - 6 \frac{\alpha}{c^2} S^{(3)} S^{(3)}_{\nu} + \beta S^{(3)}_{\nu \nu \nu} \rb_{\nu}. \label{Q3exp}
\end{align}
We define the averaging in this region as $\displaystyle \lim_{\chi \to \infty} \frac{1}{\chi - x_b} \int_{x_b}^{\chi} \dots \ dx$. Averaging at constant $\xi$ and $\nu$ and integrating with respect to the appropriate characteristic variable, we obtain two KdV equations of the form
\begin{align}
T^{(3)}_{X} - 6 T^{(3)} T^{(3)}_{\xi} + T^{(3)}_{\xi \xi \xi} &= 0, \label{T3eq} \\
S^{(3)}_{X} - 6 \frac{\alpha}{c^2} S^{(3)} S^{(3)}_{\nu} + \beta S^{(3)}_{\nu \nu \nu} &= 0. \label{S3eq}
\end{align}
Substituting the results for \eqref{T3eq} into \eqref{P3exp} and integrating with respect to the characteristic variables, we obtain
\begin{equation}
P^{(3)} = \phi_1^{(3)} \lb \xi, X \rb + \psi_1^{(3)} \lb \eta, X \rb,
\label{P3eq}
\end{equation}
where $\phi_1^{(3)}$, $\psi_1^{(3)}$ are arbitrary functions. The second radiation condition states that if the incident wave is right-propagating, then there should be no left-propagating waves in the transmitted wave field. \cite{KS, KT} Thus, $\psi_1^{(3)} = 0$.

Similarly, substituting \eqref{S3eq} into \eqref{Q3exp} and integrating with respect to the characteristic variables, we obtain
\begin{equation}
Q^{(3)} = \phi_2^{(3)} \lb \nu, X \rb + \psi_2^{(3)} \lb \zeta, X \rb,
\label{Q3eq}
\end{equation}
where $\phi_2^{(3)}$, $\psi_2^{(3)}$ are arbitrary functions. By the same argument as above $\psi_2^{(3)} = 0$. 

\subsection{Matching at boundaries: continuity conditions}
\label{sec:BC}
In order to find `initial conditions' for the equations derived by the averaging, we collect the expressions for the weakly nonlinear solutions and substitute them into the continuity conditions (\ref{cont_a}) - (\ref{cont2_b}).

We first consider the continuity conditions for displacements for the time interval when the waves have not yet reached the third region. The displacement at negative infinity is assumed to be constant. Differentiating the continuity conditions (\ref{cont_a}) with respect to time at $x = x_a$, and recalling that $f^{(i)} = u^{(i)}_x$, $g^{(i)} = w^{(i)}_x$, we obtain the following conditions in terms of the strain rates:
\begin{align}
\int_{-\infty}^{x_a} f^{(1)}_t \dd{x} &= -\int_{x_a}^{x_b} f^{(2)}_t \dd{x},  \label{udiff1} \\
\int_{-\infty}^{x_a} g^{(1)}_t \dd{x} &= -\int_{x_a}^{x_b} g^{(2)}_t \dd{x}. \label{wdiff1}
\end{align}
Substituting the weakly nonlinear solutions obtained in Section \ref{sec:WNL} into \eqref{udiff1} and noting that the reflected waves $R^{(2)}$ and $G^{(2)}$ in the second section are not yet present, we obtain at leading order
\begin{equation}
\int_{-\infty}^{x_a} (I_{\xi} - R^{(1)}_{\eta}) \dd{x} = - \int_{x_a}^{x_b} T^{(2)}_{\xi} \dd{x}.
\label{uint1}
\end{equation}
We can integrate to obtain an expression at $x=x_a$ by noting that integration with respect to $x$ can be reduced to integration with respect to the characteristic variable, as $x$ appears linearly in the expressions for the characteristic variables. By the assumption that the strain waves are localised in the first two regions, when evaluated at either $x=-\infty$ or $x=x_b$ the expression will be zero. Therefore, from \eqref{uint1} we obtain
\begin{equation}
I|_{x=x_a} - R^{(1)}|_{x=x_a} = T^{(2)}|_{x=x_a}.
\label{ucont1xa}
\end{equation}
Similarly,  from \eqref{wdiff1} we obtain
\begin{equation}
I|_{x=x_a} - G^{(1)}|_{x=x_a} = c S^{(2)}|_{x=x_a}.
\label{wcont1xa}
\end{equation}

Next, we consider the continuity conditions for displacements for the time interval when the localised strain waves are present in all three regions, but the waves reflected from the boundary $x = x_b$ between the second and the third region have not yet reached the boundary $x = x_a$ between the first and the second region. The displacement at positive infinity is assumed to be equal to zero (the waves propagate into the unperturbed media). Differentiating the continuity conditions (\ref{cont_b}) with respect to time at $x = x_b$, and recalling that $f^{(i)} = u^{(i)}_x$, $g^{(i)} = w^{(i)}_x$, we obtain the following conditions in terms of the strain rates:
\begin{align}
\int_{x_a}^{x_b} f^{(2)}_t \dd{x} &= -\int_{x_b}^{\infty} f^{(3)}_t \dd{x}, \label{udiff2} \\
\int_{x_a}^{x_b} g^{(2)}_t \dd{x} &= -\int_{x_b}^{\infty} g^{(3)}_t \dd{x}. \label{wdiff2}
\end{align}

Then, similarly to the previous considerations, we obtain
\begin{align}
T^{(2)}|_{x=x_b} - R^{(2)}|_{x=x_b} &= T^{(3)}|_{x=x_b}, \label{ucont1x0} \\
c S^{(2)}|_{x=x_b} - c G^{(2)}|_{x=x_b} &= c S^{(3)}|_{x=x_b}. \label{wcont1x0}
\end{align}
We now make use of the continuity conditions for normal stress and, substituting the relevant weakly nonlinear solution into \eqref{cont2_a} (noting that $u_x^{(i)} = f^{(i)}$ and $w_x^{(i)} = g^{(i)}$) we obtain to leading order
\begin{align}
I|_{x=x_a} + R^{(1)}|_{x=x_a} &= T^{(2)}|_{x=x_a} \label{ucont2xa} \\
I|_{x=x_a} + G^{(1)}|_{x=x_a} &= c^2 S^{(2)}|_{x=x_a}. \label{wcont2xa}
\end{align}
Similarly, substituting the relevant weakly nonlinear solutions into \eqref{cont2_b} we have, to leading order,
\begin{align}
T^{(2)}|_{x=x_b} + R^{(2)}|_{x=x_b} &= T^{(3)}|_{x=x_b} \label{ucont2x0} \\
c^2 S^{(2)}|_{x=x_b} + c^2 G^{(2)}|_{x=x_b} &= c^2 S^{(3)}|_{x=x_b}. \label{wcont2x0}
\end{align}
We can now find `initial conditions' for the systems describing transmitted and reflected waves in each section of the bar, expressed in terms of the given incident wave. For the top layer we have
\begin{align*}
&R^{(1)}|_{x=x_a} = C_R^{(1)} I|_{x=x_a}, &&T^{(2)}|_{x=x_a} = C_T^{(1)} I|_{x=x_a}, \\ 
&R^{(2)}|_{x=x_b} = C_R^{(2)} T^{(2)}|_{x=x_b}, &&T^{(3)}|_{x=x_b} = C_T^{(2)} T^{(2)}|_{x=x_b}, 
\end{align*}
where $C_R^{(i)} = 0$ and $C_T^{(i)} = 1$ for all $i$.

Similarly for the bottom layer we have 
\begin{align*}
&G^{(1)}|_{x=x_a} = D_R^{(1)} I|_{x=x_a}, &&S^{(2)}|_{x=x_a} = D_T^{(1)} I|_{x=x_a}, \\ 
&G^{(2)}|_{x=x_b} = D_R^{(2)} S^{(2)}|_{x=x_b}, &&S^{(3)}|_{x=x_b} = D_T^{(2)} S^{(2)}|_{x=x_b}, 
\end{align*}
where
${\displaystyle D_R^{(1)} = \frac{c - 1}{c + 1}, D_T^{(1)} = \frac{2}{c \lb 1 + c \rb}, 
D_R^{(2)} = 0, D_T^{(2)} = 1.} $
These coefficients agree with previous work  for a perfectly bonded waveguide  \cite{KS, KT} and are intuitive, as we would expect a wave to be, to leading order, wholly transmitted when travelling along a layer of the same material. If the value of $c$ varies between sections of the bar i.e. the material in a single layer varies across the bar, then the coefficients should be calculated using the respective values of $c$. 

We note that the functions which remained undefined in the higher order corrections can be found by considering higher order terms in the equations of motion and the continuity conditions, similarly to the solution of the initial-value problems. \cite{KM, KMP14} However, this is beyond the scope of our present paper.

\section{Numerical modelling}
\label{sec:InfDelam}
In a case study we compare the results of the semi-analytical numerical modelling, based on the leading order terms of the weakly nonlinear solution of the previous section, with the results of direct numerical simulations for the problem (\ref{syshomog}) - (\ref{cont2_b}).  We also compare numerical results with theoretical predictions for the amplitude of the lead soliton in the delaminated region, made using the IST. We solve the original Boussinesq equations using the finite-difference method described in Appendix \ref{sec:FDM}, 
and the weakly nonlinear solution derived in Section \ref{sec:WNL} using the pseudospectral method described in Appendix \ref{sec:PS}. 
For the finite-difference method, in all cases, we use step sizes of $\Delta x = \Delta t = 0.01$ and, for the pseudospectral method, we use $\Delta \xi = 0.3$ (the same step size is used for all  characteristic variables) and $\Delta X = 0.001$. In all cases, we assume $\alpha = 1.05$, $\beta = 1.05$, $c = 1 + \epsilon/2$ and $\epsilon = 0.05$. 

We note that, similarly to the single Ostrovsky equation, the coupled Ostrovsky equations (\ref{T2eq}) - (\ref{S2eq}) and (\ref{R2eq}) - (\ref{G2eq}) imply zero mass constraints:
\begin{equation*}
\int_{-\infty}^{\infty} (T^{(2)} - S^{(2)}) d \xi = 0 \ \mbox{and} \ \int_{-\infty}^{\infty} (R^{(2)} - G^{(2)}) d \eta = 0.
\end{equation*}
Therefore we first use initial conditions for the incident strain solitary wave which include a pedestal term, \cite{KMP14} to guarantee zero mass, and then show that for this class of problems, one can also work with initial conditions in the form of pure strain solitary waves, without the pedestal terms. Indeed, in the latter case the zero mass constraints are still approximately satisfied, by the nature of the solutions of the problem, which we established in the direct numerical simulations using the finite-difference method.  

Thus, we use the following initial conditions for the displacements in  \eqref{syshomog} (corresponding localised initial conditions for the strains in (\ref{syshomogdiff}) are given by the derivatives of these functions)
{\small
\begin{align}
u \lb x, 0 \rb &= A\ \lsq \tanhn{}{\frac{x}{\Lambda}} - 1 \rsq \notag \\
 &- \Gamma \lsq \tanhn{}{\frac{x + x_0}{\Lambda S}} + \tanhn{}{\frac{x - x_0}{\Lambda S}} - 2 \rsq, \notag \\
u \lb x, \kappa \rb &= A\ \lsq \tanhn{}{\frac{x - \kappa v}{\Lambda}} - 1 \rsq  \nonumber \\
&- \Gamma \lsq \tanhn{}{\frac{x + x_0 - \kappa v}{\Lambda S}}  + \tanhn{}{\frac{x - x_0 - \kappa v}{\Lambda S}} - 2 \rsq,
\label{uIC}
\end{align}
}where we have $A = - \frac{v \sqrt{v^2 - 1}}{\sqrt{2 \epsilon}}$, $\Lambda = \frac{ 2 \sqrt{2 \epsilon} v }{\sqrt{ v^2 - 1}}$ and
{\small
\begin{equation*}
\Gamma =  \frac{\sigma A\ \tanhn{}{\frac{L}{\Lambda}}}{\tanhn{}{\frac{L + x_0}{\Lambda S}} + \tanhn{}{\frac{L - x_0}{\Lambda S}}}, 
\end{equation*}
}$\sigma$ can be zero or one, $2L$ is the length of the domain used in the weakly nonlinear modelling, $x_0$ is an arbitrary phase shift, $\kappa = \Delta t$, and the corresponding strain has zero mass  (for $\sigma = 1$). The constant of integration is chosen so that the waves propagate into an unperturbed medium. 
The amplitude of the pedestal for the corresponding strains can be reduced by increasing the value of $S$. In all cases discussed here, $S = 10$ and we set $x_0 = 0$. For $w(x,0)$ and $w(x,\kappa)$ we use the same expressions. If the initial condition was not given by an explicit analytic function, then we could deduce the second initial condition for the scheme by taking the forward difference approximation (simulations have shown that either case is sufficiently accurate).

For the weakly nonlinear solutions we take the exact solitary wave solution of the equation \eqref{Ieq} governing the incident wave, with the same pedestal term (differentiated with respect to $x$) as used in \eqref{uIC},
{\small
\begin{align}
I \lb \xi, 0 \rb =& \tilde A\ \sechn{2}{\frac{\xi}{\tilde \Lambda}} \nonumber \\
 -& \frac{ \tilde \Gamma}{S} \lsq \sechn{2}{\frac{\xi + x_0}{\tilde \Lambda S}} + \sechn{2}{\frac{\xi - x_0}{\tilde \Lambda S}} \rsq, \qquad
\label{IIC}
\end{align}
}where $\tilde A = -\frac{v_1}{2}$, $\tilde \Lambda = \frac{2}{\sqrt{v_1}}$ and 
{\small
\begin{equation*}
\tilde \Gamma = \frac{\sigma \tilde A\ \tanhn{}{\frac{L}{\tilde \Lambda }}}{\tanhn{}{\frac{L + x_0}{\tilde \Lambda S}} + \tanhn{}{\frac{L - x_0}{\tilde \Lambda S}}},
\end{equation*}
}where $v_1$ is related to $v$ by the approximate relation $v = 1 + \epsilon v_1 + \O{\epsilon^2}$. For the initial conditions in other sections of the bi-layer, we make use of the relations in Section \ref{sec:BC} to obtain the initial conditions in terms of \eqref{IIC}.

\subsection{Solitons in the delaminated section}
\label{sec:Scattering}
To obtain quantitative predictions for parameters of solitons in the delaminated section we use the IST 
applied to the KdV equations (\ref{T3eq}) and (\ref{S3eq}) derived in Section \ref{sec:R3}. 
Firstly we recall that the solution of an initial-value problem for the KdV equation
\begin{equation}
U_{\tau} - 6 U U_{\chi} + U_{\chi \chi \chi} = 0, \quad U |_{\tau = 0} = U_0(\chi)
\label{KdVCanon}
\end{equation}
on the inifinite line, for a  sufficiently rapidly decaying initial condition $U_0(\chi)$,
is related to the spectral problem for the Schr\"{o}dinger equation
\begin{equation}
\Psi_{\chi \chi} + \lsq \lambda - U_0 \lb \chi \rb \rsq \Psi = 0, 
\label{SchroEq}
\end{equation}
where $\lambda$ is the spectral parameter. \cite{GGKM} In particular, parameters of solitons are defined by the discrete spectrum of the equation (\ref{SchroEq}). In our previous studies of the scattering of an incident solitary wave in the delaminated area of the perfectly bonded waveguide, the discrete spectrum could be found analytically. \cite{KS, KT} However, in the present study, we are dealing with the scattering of radiating solitary waves, generated in the  two-layered bar with  soft bonding, and scattered in the delaminated region of the bar. Therefore, we have to find the spectrum of the Schr\"{o}dinger equation numerically.

To achieve this we implement a shooting method. \cite{SB} We consider the potential $U_0(\chi)$ for the Schr\"{o}dinger equation, which is the initial condition in the KdV problem
\eqref{KdVCanon}. It is well known that the discrete spectrum is bounded by the minimum of the initial condition (negative value) and zero. \cite{LL} Since the potential $U_0(\chi)$ is localised, the eigenfunctions have the asymptotic behaviour 
\begin{equation}
	\Psi \lb \chi \rb = 
	\begin{cases}
		e^{r \chi}, & \chi \rightarrow -\infty, \\
		e^{-r \chi}, & \chi \rightarrow \infty,
	\end{cases}
	\label{EigFunc}
\end{equation}
where $\lambda = - r^2$. We rewrite the Schr\"{o}dinger equation (\ref{SchroEq}) in the form $\Psi_{\chi} = \Phi$, $\Phi_{\chi} = \lsq U_0 \lb \chi \rb - \lambda \rsq \Psi$, and solve this system from the boundary $\chi = a$ to $\chi = b$. We normalise the solution by setting $\Psi(a) = 1$ and $\Phi(a) = r$. The system is then solved using the standard Runge-Kutta 4$^{\mathrm{th}}$ order method. The ratio of the values of these two functions  is  tested at the right boundary against the relation $\Phi(b)/ \Psi(b) = - r$ to determine if $r$ is an eigenvalue.
We start with the least possible value for an eigenvalue (the minimum of the initial condition) 
and iterate to zero in sufficiently small steps in order to determine the eigenvalues to the desired accuracy. In our present study we consider the cases when in each layer there is only one soliton in the delaminated region. We use the method described here to determine the parameters of this soliton (in each layer), and compare with the solitons emerging in our modelling. In other settings, multiple solitons can be generated by a single incident soliton. \cite{KS, KT}

\subsection{Delamination of semi-infinite length}
\label{sec:SemiFinDelam}
\begin{figure*}
\includegraphics[width = 0.9\textwidth, trim = 0cm 0.5cm 0cm 1cm]{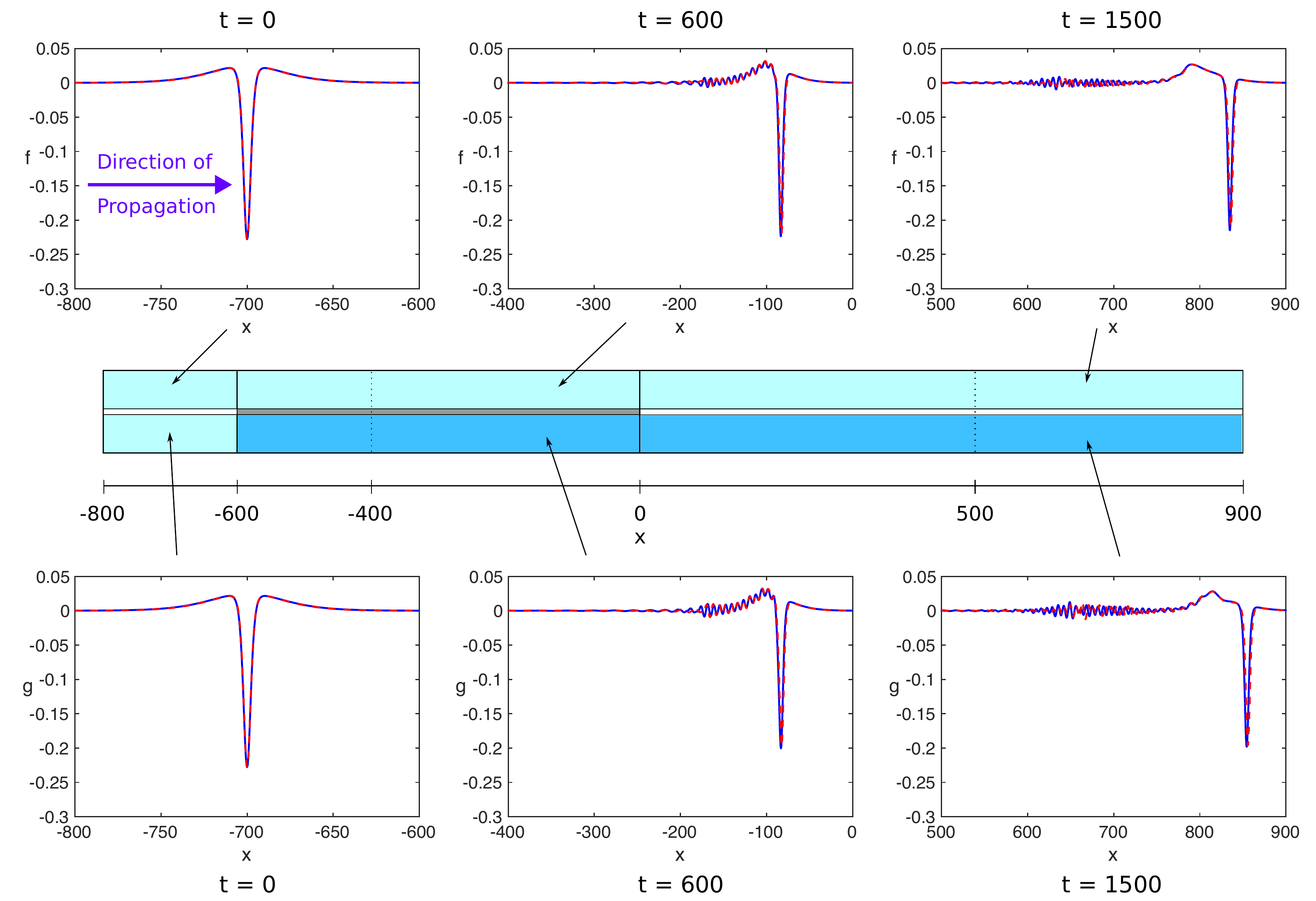}
\caption{The waves $f$ (top row) and $g$ (bottom row) in the various sections of the bi-layer, for $\alpha = \beta = 1.05$, $c = 1.025$, $\delta = \gamma = 1$, $v = 1.025$ , $\sigma = 1$ and $\epsilon = 0.05$:  direct numerical simulations (solid line) and weakly nonlinear solution (dashed line). For the finite-difference method, the full computational domain is $[-1000, 1000]$. In the pseudospectral method, $N = 16384.$}
\label{fig:RSWe05}
\end{figure*}

We first consider the bi-layer shown in Figure \ref{fig:DelamBar} and use the initial conditions (\ref{uIC}) and (\ref{IIC}) with zero mass, i.e. with $\sigma = 1$. The comparison between the two numerical approaches in this case can be seen in Figure \ref{fig:RSWe05}. A radiating solitary wave is formed in the bonded section of the bar, shown at $t=600$. When this radiating solitary wave enters the delaminated section of the bar, the soliton separates from the tail and becomes a classical soliton with dispersive radiation following behind.
The agreement between the weakly nonlinear solution and the direct finite-difference technique is good for the solitons and reasonable for the tail, with a small phase shift introduced. The agreement is improved by reducing $\epsilon$, and this has been tested for a number of values.

\begin{figure*}
\includegraphics[width = 0.9\textwidth, trim = 0cm 0.5cm 0cm 1cm]{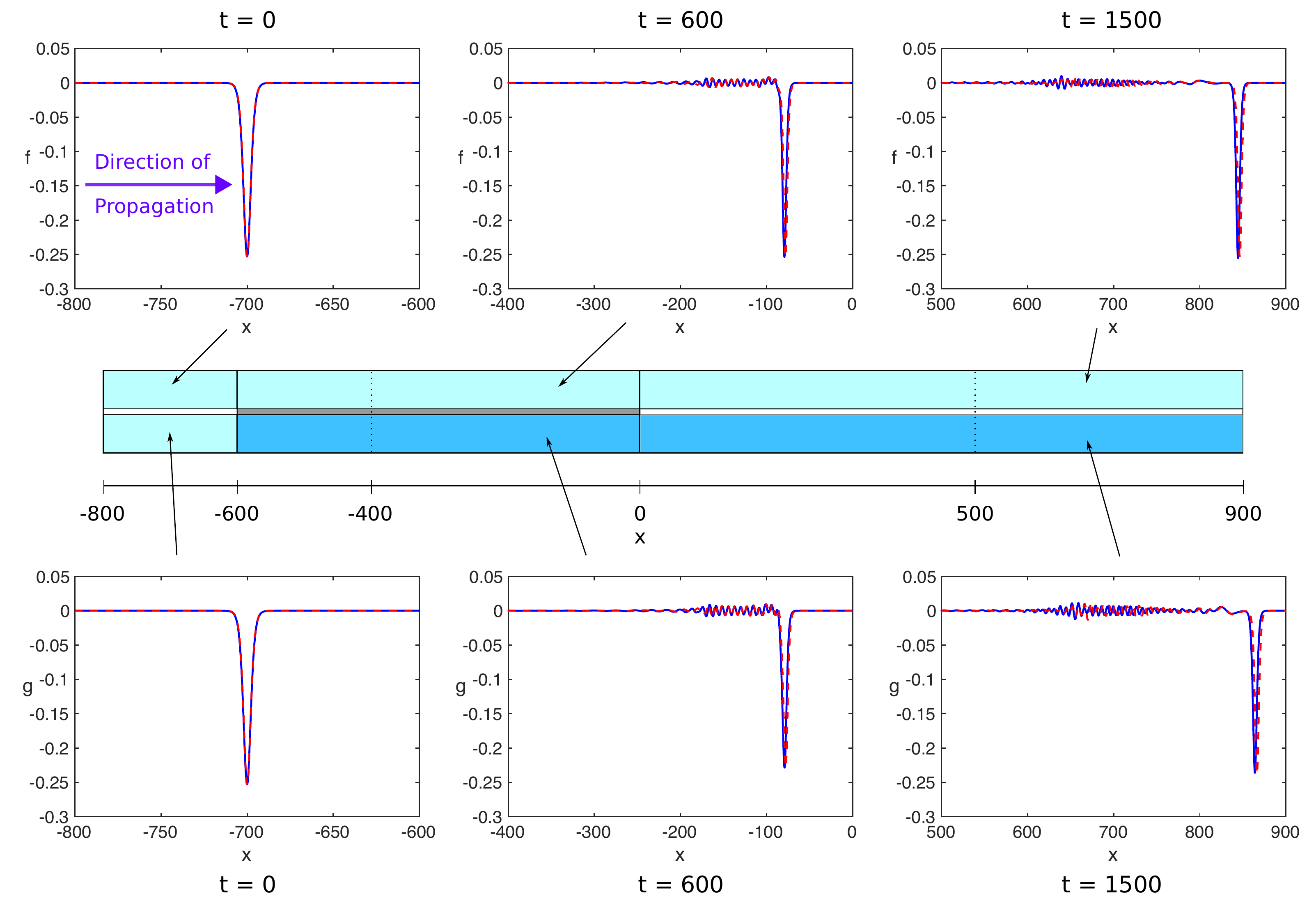}
\caption{The waves $f$ (top row) and $g$ (bottom row) in the various sections of the bi-layer,  for $\alpha = \beta = 1.05$, $c = 1.025$, $\delta = \gamma = 1$, $v = 1.025$, $\sigma = 0$  and $\epsilon = 0.05$:  direct numerical simulations (solid line) and weakly nonlinear solution (dashed line). For the finite-difference method, the full computational domain is $[-1000, 1000]$. In the pseudospectral method, $N = 16384.$}
\label{fig:RSWe05XZM}
\end{figure*}

If we take the same initial conditions with non-zero mass, i.e. $\sigma = 0$,  we obtain a similar result to
the previously discussed case, as can be seen in Figure \ref{fig:RSWe05XZM}. The radiating solitary waves generated in the two layers are close to each other, and therefore the zero mass constraints 
for the difference of the two solutions  are approximately satisfied (see Appendix \ref{sec:PS}).

We now apply the IST framework to the waves entering the delaminated section of the bar, as the behaviour of the transmitted waves in the two layers in this section is governed by two separate KdV equations. We numerically solve the scattering problem for the related Schr\"{o}dinger equation, as discussed in Section \ref{sec:Scattering}, to obtain the eigenvalues. Since there is only one discrete eigenvalue for each layer of the waveguide, the long time asymptotic behaviour of the solution of the appropriate KdV equation  consists of one soliton and dispersive radiation, which in the canonical form (\ref{KdVCanon}) is given by 
\begin{equation*}
U \sim - 2 r^2 ~\mathrm{sech}^2 \lsq r \lb \chi - 4 r^2 t - \chi_0 \rb \rsq + \text{radiation},
\end{equation*}
where $r$ is defined by the eigenvalue $\lambda = - r^2$,  and $\chi_0$ is the phase shift. 

We use the theoretical predictions to justify the numerical schemes in Appendices \ref{sec:FDM} and \ref{sec:PS}. 
In each layer, the height of the soliton found using these schemes has been  compared with the theoretical prediction using the IST, to confirm that the numerical schemes resolve the behaviour of the system correctly. The theoretical (IST) predictions and the numerical results for the height of the soliton are compared in Table \ref{tab:Schro}.
\begin{table}[!ht]
\centering
\begin{tabularx}{0.48\textwidth}{| c | Y | Y | Y |}
\hline
Regime & Layer & Numerical & Theoretical \\ \hhline{|=|=|=|=|}
$\sigma = 1$ & 1 & -0.2545 & -0.2473 \\ \hline
$\sigma = 1$ & 2 & -0.2301 & -0.2192 \\ \hline
$\sigma = 0$ & 1 & -0.2979 & -0.2979 \\ \hline
$\sigma = 0$ & 2 & -0.2680 & -0.2680 \\ \hline
\end{tabularx}
\caption{Comparison of amplitudes for solitons in the delaminated area for both layers with the predicted value using the IST, for zero mass ($\sigma = 1$) and non-zero mass ($\sigma = 0$) initial conditions.}
\label{tab:Schro}
\end{table}

In the case with zero mass initial condition, the prediction of the heights using the IST underestimates the numerical solution, as the solitons have not yet fully separated from the negative pedestal. In the case with initial condition having non-zero mass, the agreement between the theoretical predictions and the numerical results is excellent.

\subsection{Delamination of finite length}
\label{sec:FinDelam}
\begin{figure}[!ht]
\center
\includegraphics[width=0.4\textwidth, trim=0mm 0mm 0mm 15mm]{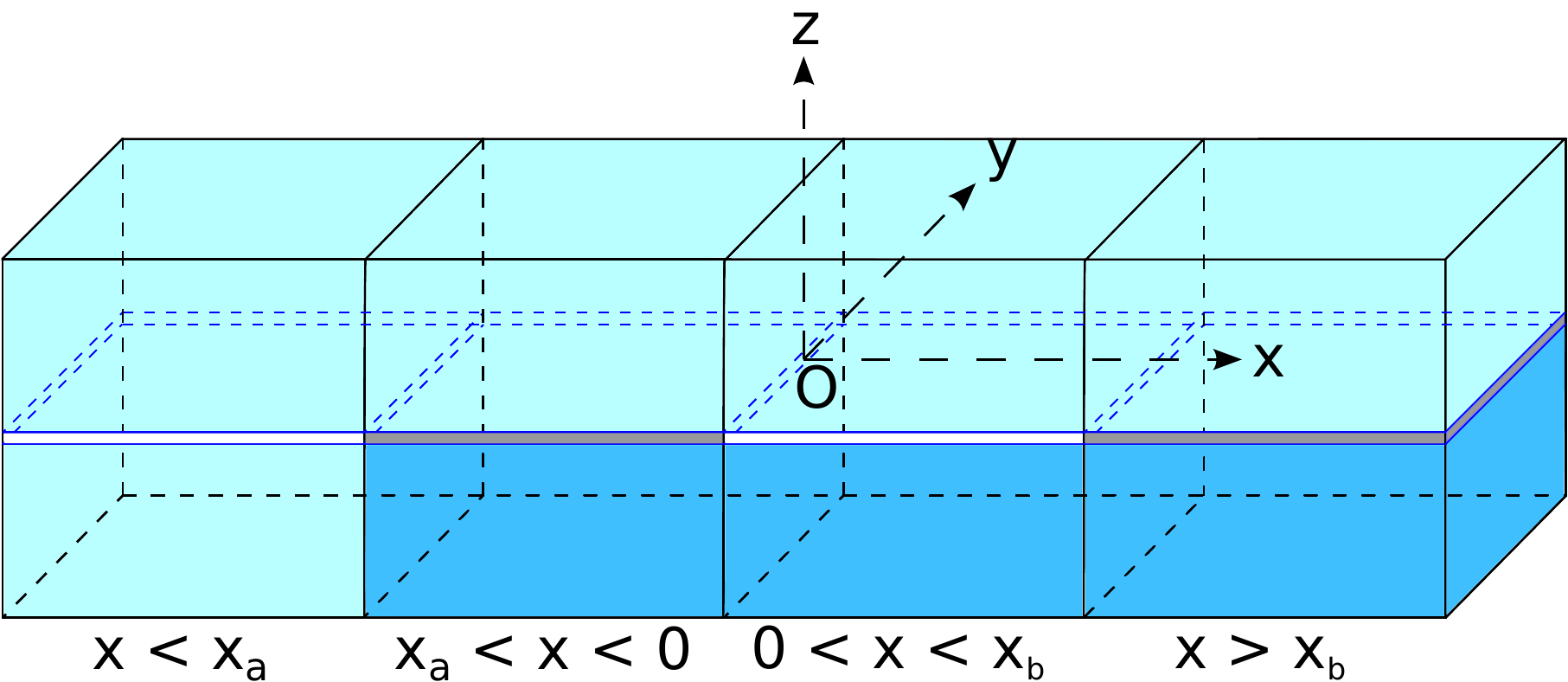}
\caption{Bi-layer with two homogeneous layers for $x < x_a$, a bonded two-layered section for $x_a < x < 0$, a delaminated section for $0 < x < x_b$ and another bonded two-layered section for $x > x_b$.}
\label{fig:CDCBar}
\end{figure}

\begin{figure*}
	\includegraphics[width = 0.9\textwidth, trim = 0cm 0.6cm 0cm 0cm]{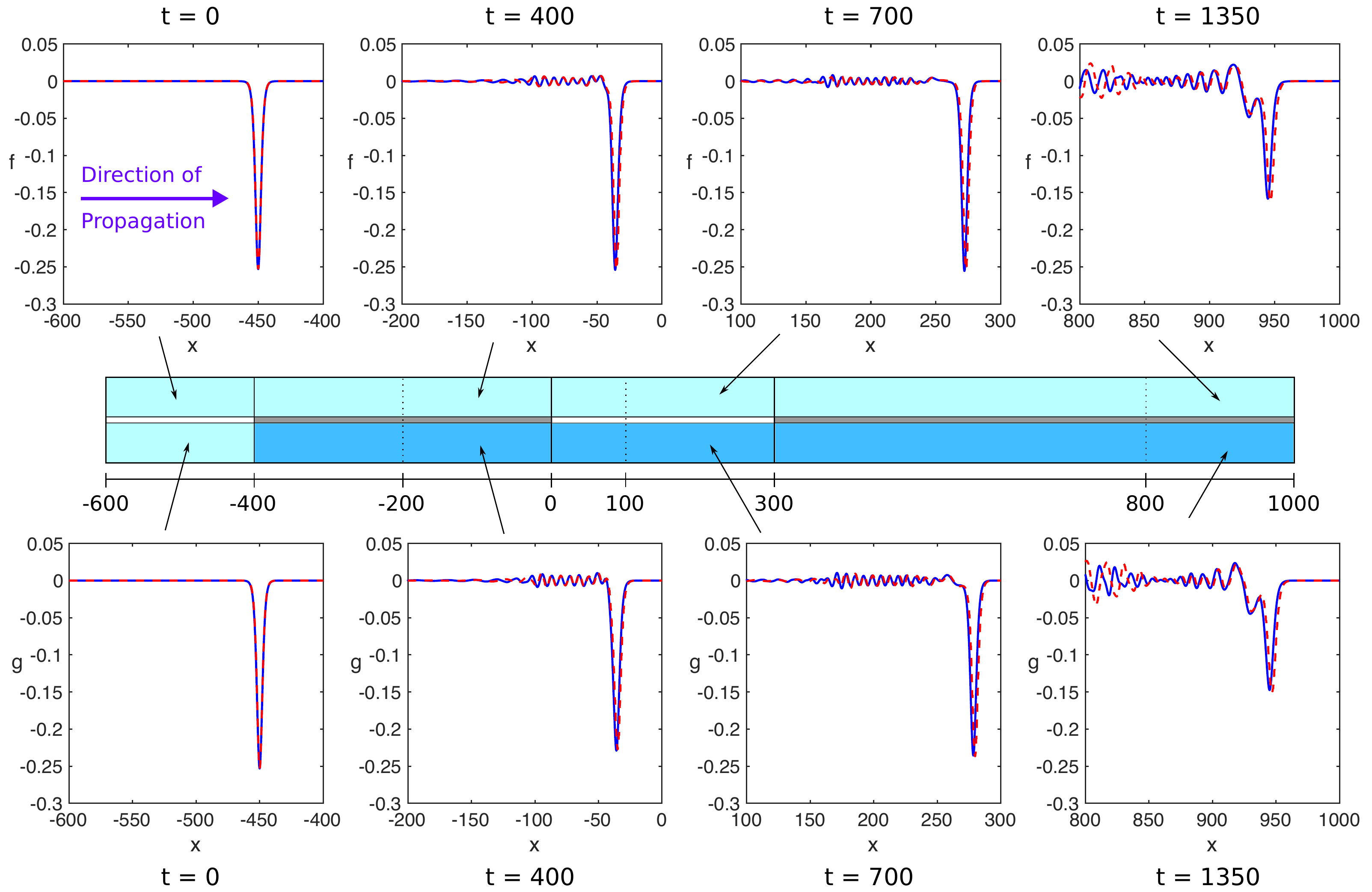}
	\caption{The waves $f$ (top row) and $g$ (bottom row) in the various sections of the bi-layer, for $\alpha = \beta = 1.05$, $c = 1.025$, $\delta = \gamma = 1$, $v = 1.025$, $\sigma = 0$ and $\epsilon = 0.05$: direct numerical simulations (solid line) and weakly nonlinear solution (dashed line). Two homogeneous layers, of the same material as the upper layer, are on the left, and the waves propagate to the right. 
 For the finite-difference method, the full computational domain is $[-600, 1000]$. In the pseudospectral method, $N = 8192.$}
	\label{fig:CDCM1L}
\end{figure*}
\begin{figure*}
	\includegraphics[width = 0.9\textwidth, trim = 0cm 0.6cm 0cm 0cm]{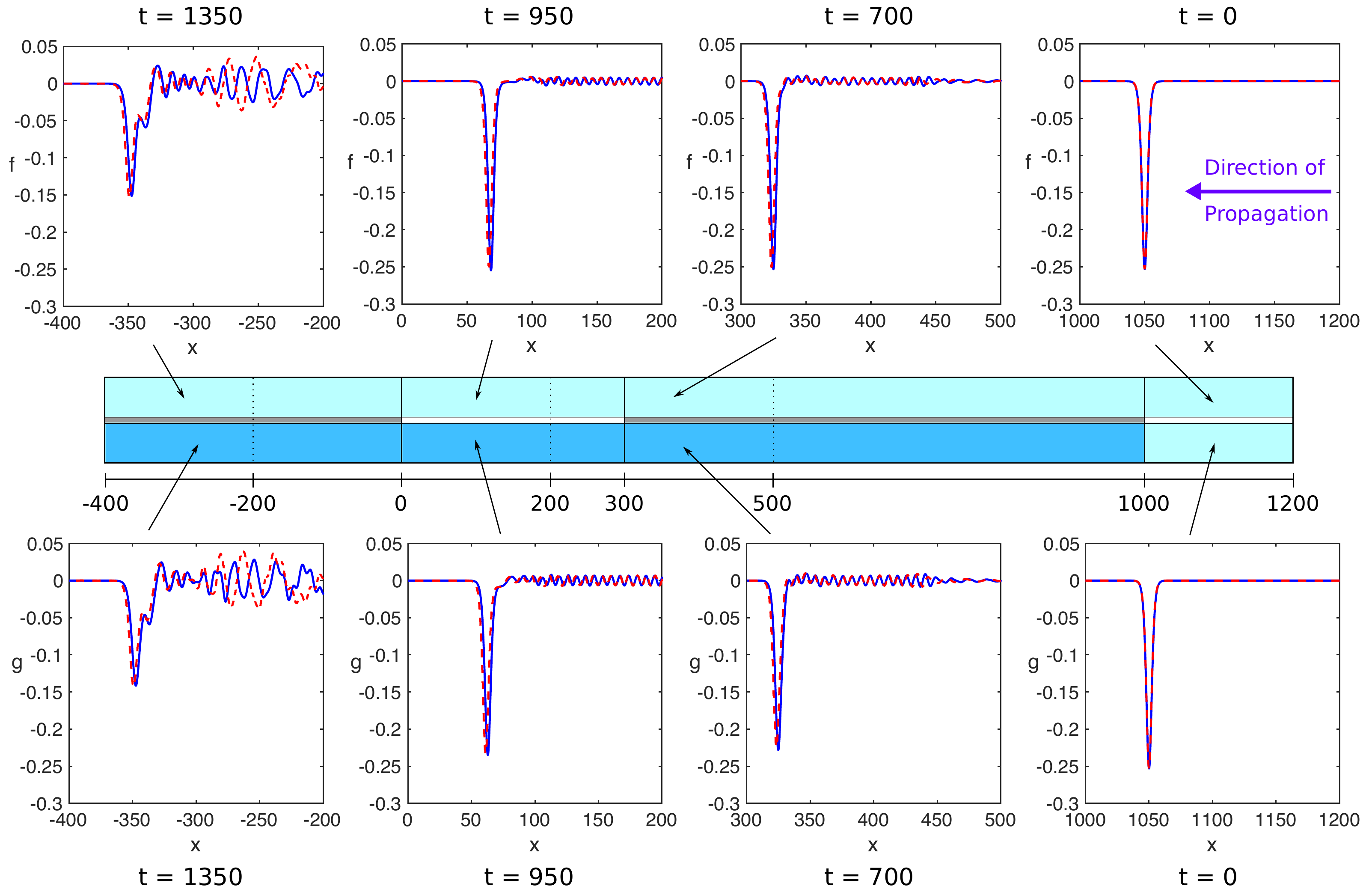}
	\caption{The waves $f$ (top row) and $g$ (bottom row) in the various sections of the bi-layer, for $\alpha = \beta = 1.05$, $c = 1.025$, $\delta = \gamma = 1$, $v = 1.025$, $\sigma = 0$ and $\epsilon = 0.05$: direct numerical simulations (solid line) and weakly nonlinear solution (dashed line). Two homogeneous layers, of the same material as the upper layer, are on the right, and the waves propagate to the left.
 For the finite-difference method, the full computational domain is $[-400, 1200]$. In the pseudospectral method, $N = 8192.$}
	\label{fig:CDCM1R}
\end{figure*}

Let us now pose a question. Is it possible to determine if there is a delamination in some part of the bar, between two bonded regions? A graphical representation of this structure is shown in Figure \ref{fig:CDCBar}, and all considerations of Section \ref{sec:WNL} are extended to this situation. We know that transmitted waves will propagate in the delaminated area  with speeds close to the characteristic speeds of the linear waves, and therefore the time it will take for the wave to travel through a delaminated region, of length $l$, can be estimated as $T_{i} \approx  l/c_i$, where $i$ represents the layers in the bar. Indeed, when the radiating solitary waves enter the delaminated region, as seen in 
Figure \ref{fig:CDCM1L}, the solitons propagate with speeds close to their respective characteristic speeds. When these solitons enter the second bonded region they again generate  radiating solitary waves. If the separation between the two solitons is sufficiently large when they enter the second bonded region, we see a distinctive double-humped wave of significantly reduced amplitude - a clear sign of delamination. 

However if the delamination area is shorter, then the solitons in the delaminated section will not be fully separated. In this case, the radiating waves in the second bonded region overlap and generate a new single-humped radiating solitary wave.  Irrespective of the separation, this process of creating a new radiating solitary wave is accompanied by some additional radiation, and therefore the amplitude of the new radiating solitary wave is reduced in both layers when compared to the radiating solitary wave propagating in a fully bonded waveguide, with no delamination. Furthermore, as the radiating solitary wave is not supported by the KdV equation, in the delaminated region the radiation separates from the soliton and the periodicity observed in the tail disappears as the tail transforms
into a wave packet. This feature gives another indication that delamination is present.

In order to investigate this behaviour more fully, we consider several cases with different delamination lengths, as measured in wavelengths of the solitary wave. 
The wavelength is measured using the common measure Full Width at Half Magnitude (FWHM). In this case, the FWHM of the incident soliton measures approximately 5 units. We present results for delamination of length 10, 20, 40 and 60 FWHM, and the case where there is no delamination. Note that Figure \ref{fig:CDCM1L} is for a delamination length of approximately 60 FWHM.
\begin{figure*}
	\begin{subfigure}[t]{0.33\textwidth}
		\includegraphics[height = 0.9\textwidth, width = \textwidth, trim = 0mm 0mm 0mm 0mm]{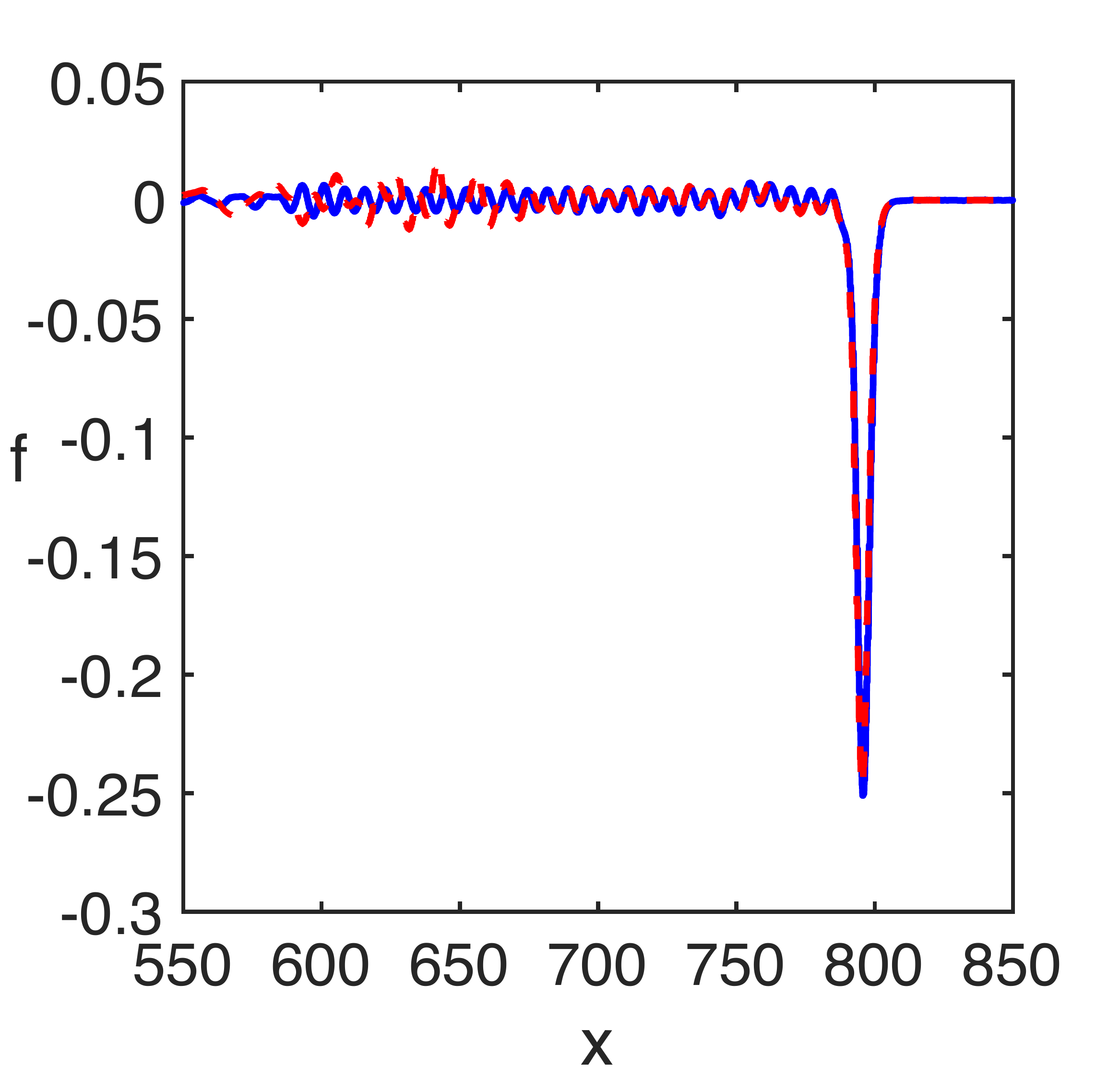}
		\caption{10 FWHM}
	\end{subfigure}
	~
	\begin{subfigure}[t]{0.33\textwidth}
		\includegraphics[height = 0.9\textwidth, width = \textwidth, trim = 0mm 0mm 0mm 0mm]{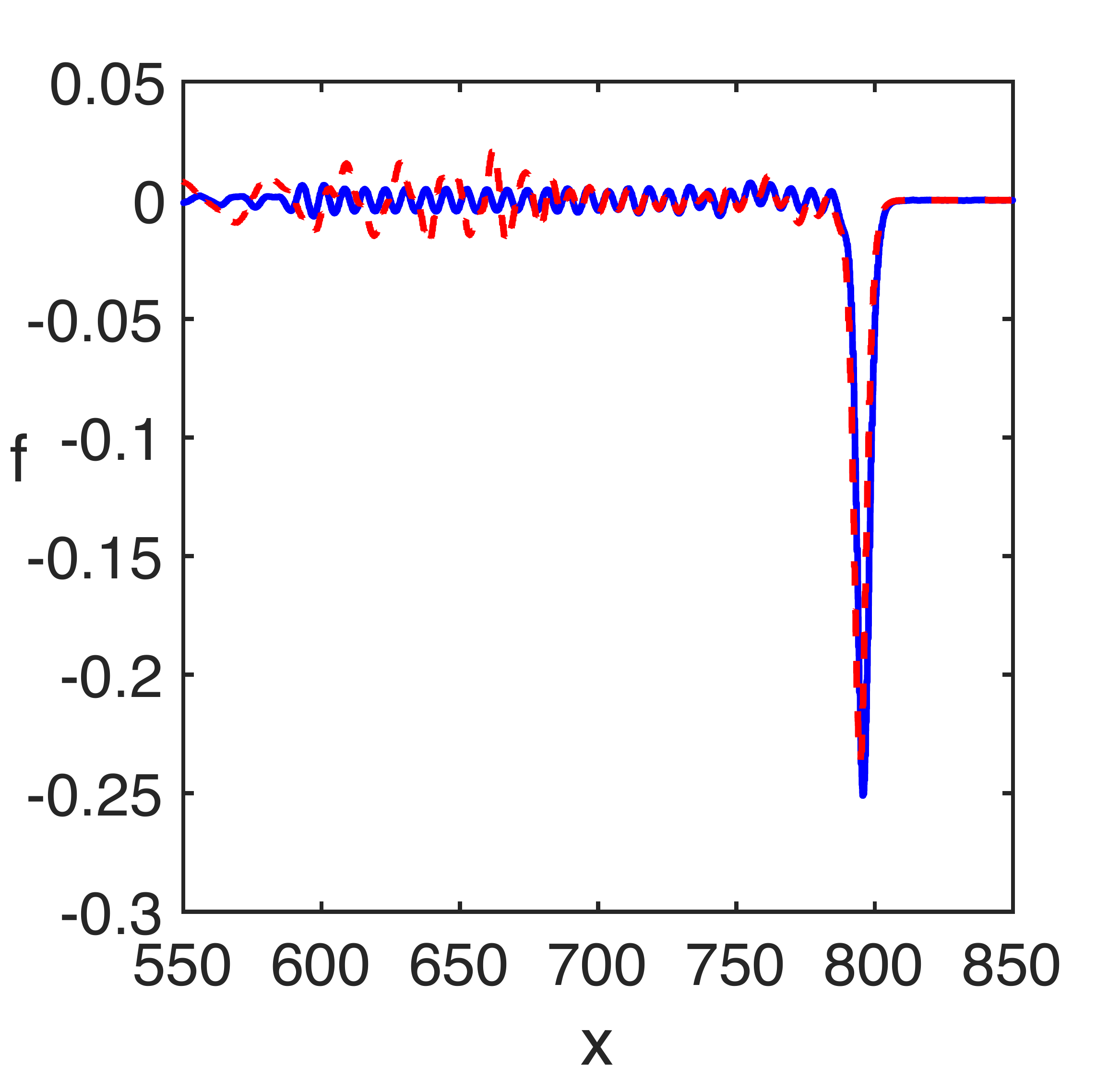}
		\caption{20 FWHM}
	\end{subfigure} \\[0.5em]
	\begin{subfigure}[t]{0.33\textwidth}
		\includegraphics[height = 0.9\textwidth, width = \textwidth, trim = 0mm 0mm 0mm 0mm]{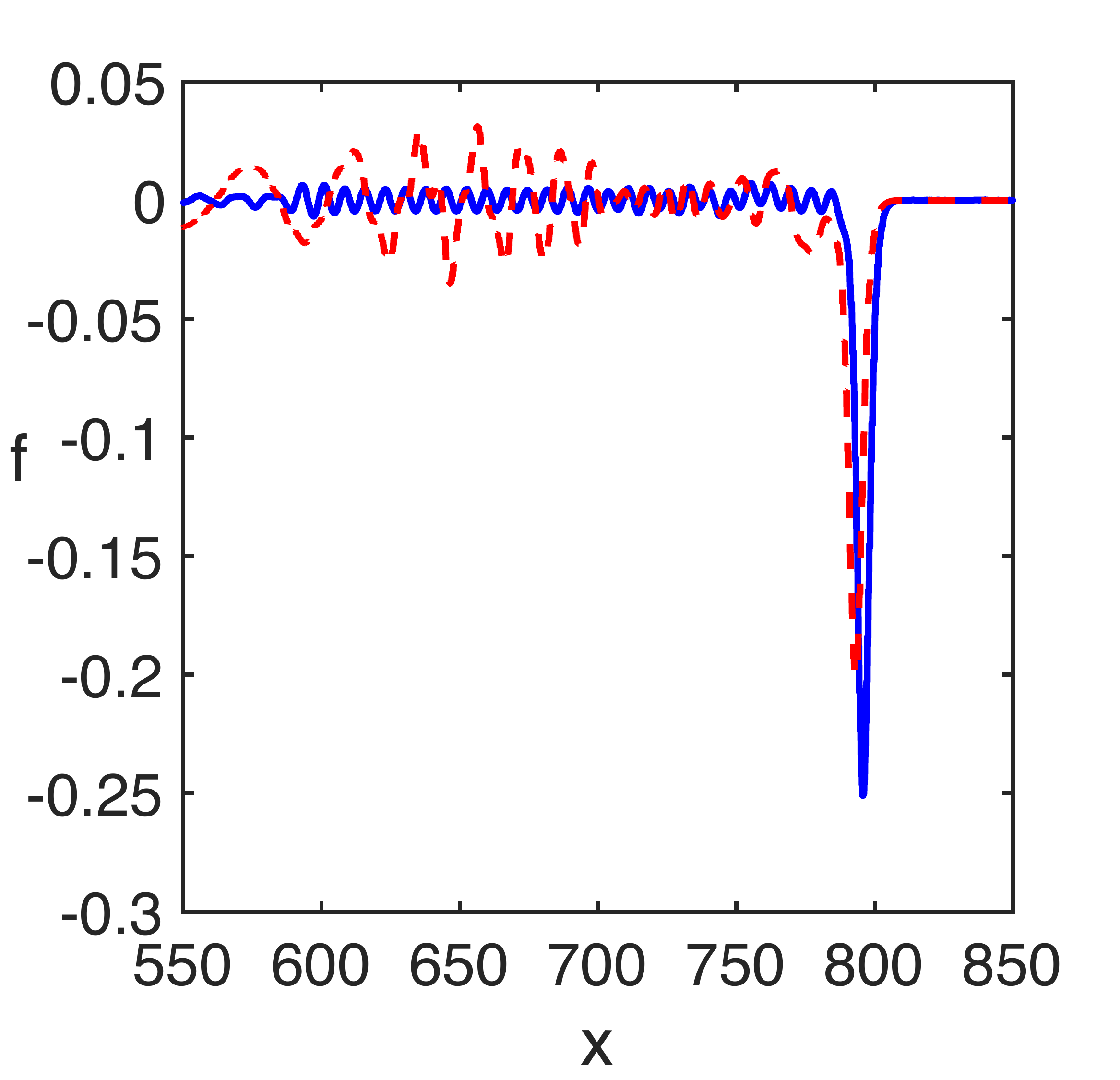}
		\caption{40 FWHM}
	\end{subfigure}
	~
	\begin{subfigure}[t]{0.33\textwidth}
		\includegraphics[height = 0.9\textwidth, width = \textwidth, trim = 0mm 0mm 0mm 0mm]{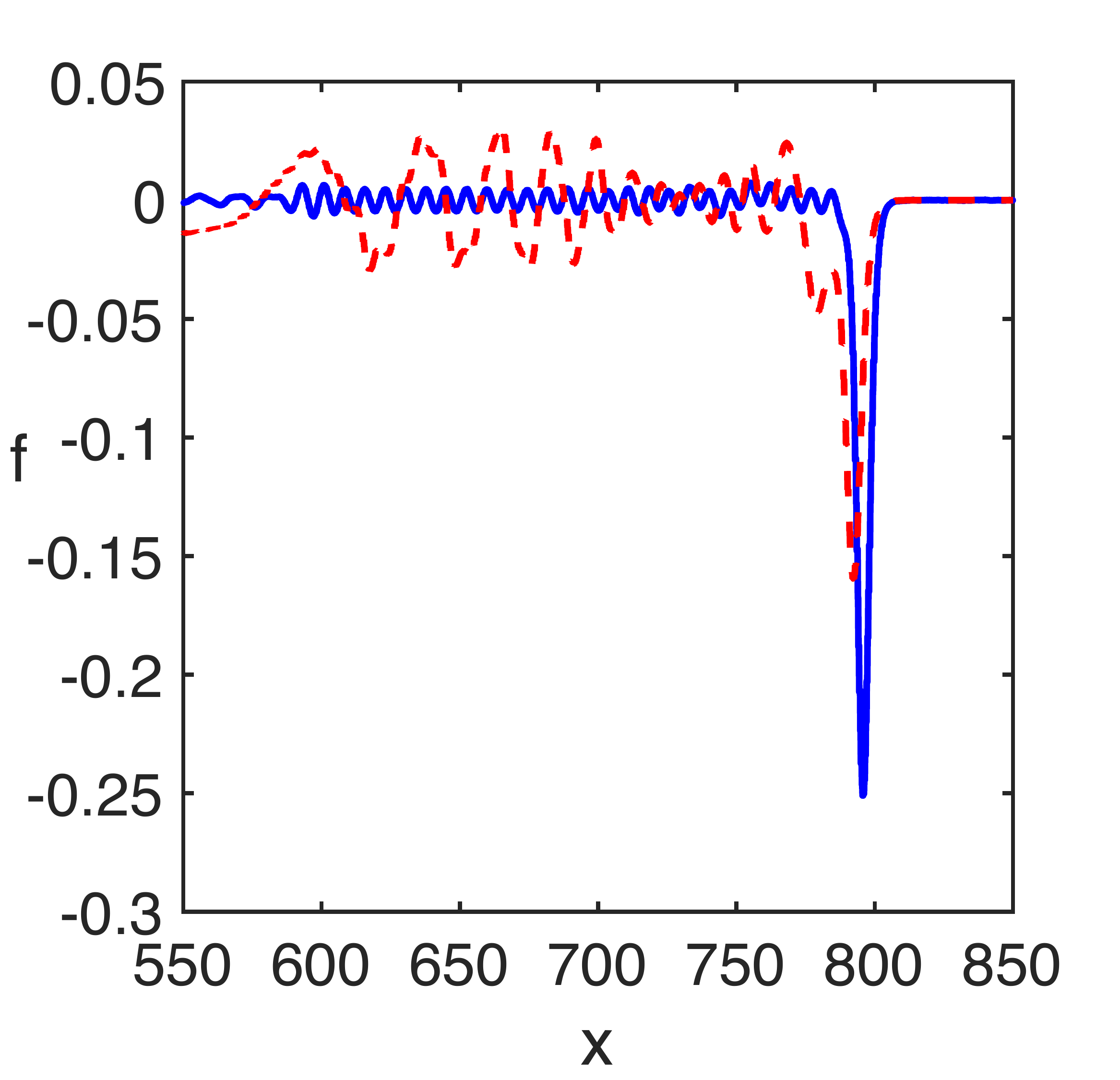}
		\caption{60 FWHM}
	\end{subfigure}
	\caption{A comparison between the case without delamination (solid lines) and with delamination (dashed lines) of differing lengths, measured in FWHM of the incident soliton. The model is the same as that used in Figure \ref{fig:CDCM1L} with the same parameters, and all images are for $t=1200$.}
	\label{fig:VarDelam}
\end{figure*}

We note that the results presented in Figure \ref{fig:VarDelam} are obtained using the semi-analytical method. The finite-difference method solves for two sections of the bar at a time and therefore 
we must wait until the wave and its tail are fully contained in the region before moving the calculation domain. However, for a shorter delamination i.e. 20 FWHM or less, the leading wave front will reach the boundary of the calculation domain before the tail has fully entered this region. Therefore, the wave will either reflect and interfere with our solution, or we will lose part of the tail when we move the calculation domain. This is a natural limitation for the use of the finite-difference method in its present form. This could be remedied by solving for all sections of the bar simultaneously, however this will be much more expensive.

We see from Figure \ref{fig:VarDelam} that there are some key differences between the model without delamination and the model with 
delamination. We only show the waves in the `top' layer as the waves in the `bottom' layer are similar.

\begin{figure}
\includegraphics[width=0.4\textwidth, trim = 15mm 0mm 15mm 0mm]{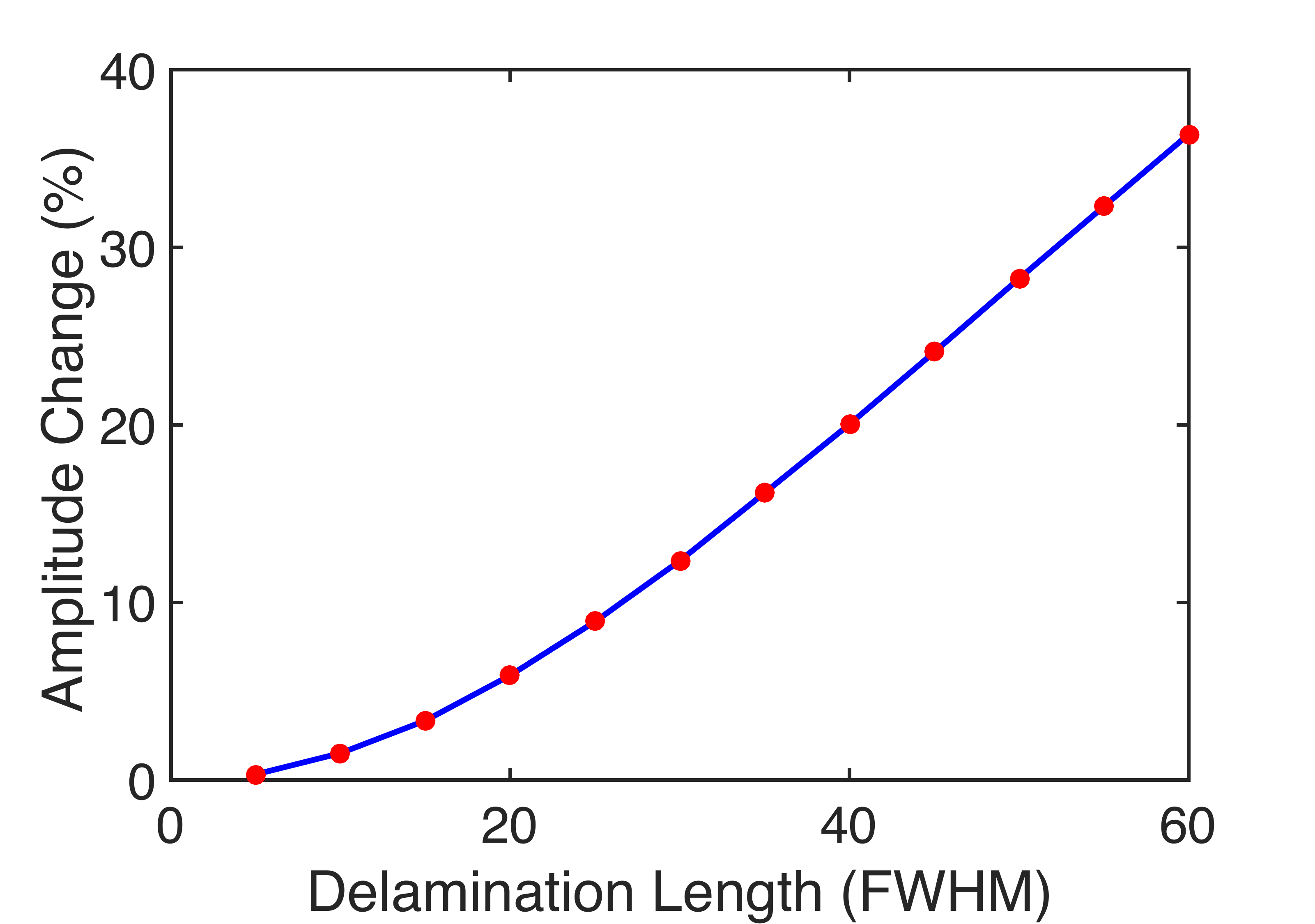}
\caption{The percentage decrease in amplitude for a given delamination measured in FWHM.}
\label{fig:Amp}
\vspace{-2em}
\end{figure}

Firstly, as the length of delamination increases, the amplitude of the radiating solitary wave created in the second bonded region is reduced. This can be explained by the fact that 
the waves in the delaminated section of the bar travel at different speeds in each layer and 
will be incident on the second bonded region at different times, so the energy exchange between layers results in the generation of a radiating solitary wave of reduced amplitude. A graph of the amplitudes against the delamination length, in FWHM, is presented in Figure \ref{fig:Amp}. We can clearly see that after an initial growth period (up to 20 FWHM), the dependence is close to linear.
The presence of the double-humped solution, as seen in the image for 60 FWHM, clearly identifies a delamination. Further numerical experiments have shown that this double-humped structure is emerging at around 45-50 FWHM. Furthermore, the small hump behind the lead soliton in the 40 FWHM image is the start of a double-humped solution, but the second `hump' has a similar amplitude to the radiation and therefore is not distinct. 

The speed of the waves in the delaminated region is different to the bonded region, and therefore when the new radiating solitary wave is formed in the second bonded region, it will have a phase shift. 
Measuring from the minima of the waves, we calculate a phase shift of 0.2, 0.8, 2.7 and 3.6 for 10, 20, 40 and 60 FWHM respectively, growing with the delamination length as expected.

The radiating solitary wave is not a solution of the KdV equation and therefore, in the delaminated region, the radiating tail forms a wave packet, breaking the regularity of the tail region. 
This feature is again more pronounced for a larger delamination area, however it can already be clearly identified for a short delamination, such as in the case of 10 FWHM as seen in Figure \ref{fig:VarDelam}. Further experiments have identified this behaviour 
for 5 FWHM, however the amplitude is similar to the rest of the tail and therefore this is difficult to identify visually. 

We summarise these observations as follows. 
For very short delamination areas i.e. 5 FWHM,  these differences are negligible and suggests that the soliton does not take care of delaminations shorter than a threshold value. For delamination areas that are greater than 5 FWHM, we can use our observations above for the amplitude reduction and phase shift to identify the presence of delamination. Modifying the FWHM value of the incident soliton can help identify shorter delaminations.

\subsection{Further experiments}
From a physical standpoint, we would like to test a given structure in as many ways as possible to obtain all possible information. Let us assume that we have a structure such as that in Figure \ref{fig:CDCBar} but with the two homogeneous layers  removed. Given this structure, there are four natural tests that we can conduct: with two homogeneous layers of either the same material as the top or bottom layers, and attaching the homogeneous layers to either the left-hand side or right-hand side of the structure, with the waves propagating to the right or to the left, respectively. Examples using the same material as the top layer are shown in Figure \ref{fig:CDCM1L} and Figure \ref{fig:CDCM1R} for the homogeneous layers being attached on the left-hand side and right-hand side respectively. We discuss all results here but omit the other cases for brevity.

Firstly we note that the double-humped structure is present in the second bonded region in all cases, confirming that each test identifies the presence of a sufficiently long delamination area, from our observations from Section \ref{sec:FinDelam}. There is a small phase shift between the results for the different materials in the homogeneous layers, arising from the higher characteristic speed of the material of the bottom layer.

We also note that, for the homogeneous layers being present on the right, the first bonded region is longer and this leads to a longer radiating tail. This tail becomes a wave packet in the delaminated region and we observe that 
the larger amplitude wave packet is closer to the double-humped structure for the case where the tail is longer, i.e. when the homogeneous section is on the right-hand side. 
Indeed, the length of the bonded region after the delamination is shorter in this case and therefore we would expect the wave packet to be closer to the leading wave. This gives us an indication of where the delamination is present in the bi-layer, i.e. if the radiation wave packet is closer to the leading wave when sending the waves from the right, then the delamination is closer to the left-hand side of the structure, and vice versa.

Another useful feature is that the generated wave is of a larger amplitude in the case when the homogeneous layers are of the same material as the bottom layer (with a larger characteristic speed), and therefore the FWHM measure is smaller. 
In addition, the amplitude difference to the case with no delamination is even clearer.

\section{Conclusions}
\label{sec:Conc}
In this paper we studied the scattering of a long radiating bulk strain solitary wave in a delaminated bi-layer with a soft bonding between the layers. The modelling was performed within the framework of the system of coupled regularised Boussinesq equations (\ref{fg}), which were derived to describe long nonlinear longitudinal waves in a two-layered waveguide with a soft bonding layer (`imperfect interface') from a layered lattice model with all essential degrees of freedom of a layered elastic waveguide \cite{KSZ}. For a single layer, the model leads to a `doubly dispersive equation' (DDE) \cite{Samsonov_book, Porubov_book},
earlier derived for the long longitudinal waves in a bar of rectangular cross-section using the nonlinear elasticity approach \cite{KS}. In dimensional variables, the DDE for a bar of rectangular cross-section $\sigma = 2a \times 2b$ has the form
\begin{equation}
f_{tt} - c^2 f_{xx} = \frac{\beta}{2 \rho} (f^2)_{xx} + \frac{J \nu^2}{\sigma} (f_{tt} - c_1^2 f_{xx})_{xx},
\label{DDE}
\end{equation}
where $c = \sqrt{E/\rho}, c_1 = c / \sqrt{2 (1+\nu)}, \beta = 3 E + 2 l (1 - 2 \nu)^3 + 4 m (1+\nu)^2 (1 - 2 \nu) + 6 n \nu^2, J = 4ab(a^2+b^2)/3$, $\rho$ is the density, $E$ is the Young's modulus, $\nu$ is the Poisson's ratio, while $l,m,n$ are the Murnaghan's moduli. Non-dimensionalisation, regularisation of the dispersive terms and scaling bring the equation to the form (\ref{f_intro}). 

The direct numerical modelling of this type of problem is difficult and expensive because one needs to solve several boundary value problems linked to each other via matching conditions at the boundaries. Therefore, we developed an alternative semi-analytical approach based upon the use of several matched asymptotic multiple-scale expansions and averaging with respect to the fast space variable. The developed approach is an extension of our earlier work, \cite{KT} where we considered a simple bi-layer with perfect bonding. Unlike our earlier work, the bi-layer with the soft (`imperfect') bonding does not support the usual solitary waves. They are replaced by radiating solitary waves, with a one-sided oscillatory tail, as discussed in the Introduction. We modelled the generation and subsequent scattering of these radiating solitary waves in a number of complex waveguides with and without a delamination area, as well as predicting the parameters of the lead solitons generated in the delaminated area using the IST framework for the relevant KdV equations. 

The developed direct finite-difference scheme and the scheme for the weakly nonlinear solution show good agreement in all regions of the bi-layer, with a small difference in the amplitude and minor phase shift between the results. This could be remedied by the inclusion of higher-order corrections in the weakly nonlinear scheme, similarly to initial-value problems. \cite{KM, KMP14} We also note that the direct finite-difference scheme is expensive in comparison to the weakly nonlinear scheme.

Our study has revealed key features of the behaviour of radiating solitary waves in such delaminated bi-layers, for different lengths of the delaminated area compared to the wavelength (FWHM) of the incident soliton. If the delaminated area is sufficiently long ($\ge 25$ FWHM), then there is a significant reduction in the amplitude of the transmitted radiating solitary wave ($\ge 10$ \%). In fact, the incident radiating solitary wave undergoes a complicated process of shedding a tail and propagating with slightly different speeds along the two layers in the delaminated region, followed by generation of a new radiating wave in the second bonded region.  For shorter delamination regions ($ < 25$ FWHM), the key dynamical effect manifesting the presence of a delaminated region in the structure is the appearance of a wavepacket in the regular tail of the radiating solitary wave. The waves are not sensitive to very short delamination regions, comparable to the wavelength of the incident soliton. In practice, using an admissible incident soliton with smaller wavelength (and higher amplitude), would increase the sensitivity to shorter delamination regions. If the delaminated region is longer ($\ge 45$ FWHM), the separation of solitons, propagating in two layers in the delaminated region, leads to the emergence of a double-humped radiating wave in the second bonded region. We did not show the cases with delamination areas greater than $60$ FWHM. The dynamical behaviour in these cases is simpler, leading to the emergence of two distinct radiating solitary waves in each layer of the second bonded region - a very clear sign of delamination. However, such cases are likely to be uncommon in real-world applications  because of the dissipation processes which have not been accounted for in our modelling. Typical values of elastic moduli for the PMMA (polymethylmethacrylate) and PS (polystyrene) and experimental data for solitons in layered PMMA/PS bars of $10 \times 10$ mm cross-section have been reported previously.  
\cite{Dreiden12, Dreiden14} The typical amplitude of the strain for the observed compression solitary waves is $\O{10^{-4}}$, and the soliton velocity is about $5-7\%$ greater than the linear longitudinal wave speed. \cite{SSB_2016}  It has been reported that, in PMMA and PS,  solitons can propagate to distances tens of times greater than their width without significant decay. \cite{SDS_2008, SSB_2016} 

The generation of a radiating bulk strain solitary wave and subsequent disappearance of the `ripples' in the delaminated area of a two-layered PMMA bar with the PCP (polychloroprene-rubber-based) adhesive has been observed in experiments. \cite{Dreiden12} Our numerical modelling motivates further laboratory experimentation with a wide range of materials used in practical applications.  It also paves the way for similar studies in other physical settings supporting radiating solitary waves and radiating dispersive shock waves, for example, in nonlinear optics. \cite{SG, CTMK}

\section*{Acknowledgments}
We thank Alexander Strohmaier and Greg Roddick  for useful discussions of numerical approaches to the calculation of the spectrum of the Schr\"{o}dinger equation. The authors would like to acknowledge the support of the Engineering and Physical Sciences Research Council (EPSRC). M.R.T. is supported by an EPSRC studentship. 

\titleformat{\section}{\bfseries}{\appendixname~\thesection .}{0.5em}{}
\appendix

\section{Numerical formulation for Boussinesq equations}
\label{sec:FDM}
The problem \eqref{syshomog} - \eqref{cont2_b} is treated as a set of boundary value problems (BVPs) with the continuity conditions defining the interaction between the equations. In order to determine the solution at a given time in a section of the bar, we solve by a method similar to that presented in our earlier work.\cite{KT} The formulation presented there was for two sections of a bi-layer, rather than three. While the formulation can be extended to a bi-layer with three sections, it will significantly increase the complexity of the numerical scheme and hence the computation time.

To alleviate this, we consider a modified approach where we compute the solution in two sections of the bi-layer at a time. Referring to Figure \ref{fig:DelamBar} as an example, we would first calculate for the region $x < x_a$ and $x_a < x < x_b$ with two constraints: that the generated radiating solitary wave has not reached the boundary $x=x_b$, and that the waves reflected from $x=x_a$ have not reflected from the left-hand boundary of the problem and into the domain $x_a < x < x_b$. The speed of the incident solitary wave and the radiating solitary wave will be close to the characteristic speed in each section of the bar, and therefore both constraints can be satisfied by choosing an appropriate time interval for the calculation. A similar approach is then followed for the region $x_a < x < x_b$ and $x > x_b$, where we choose an appropriate time so that the waves in this section have not reflected from the boundary $x=x_b$ and back again from the boundary $x = x_a$.

We summarise the method for, say, the second and third sections. Discretising the domain $\lsq x_a, L \rsq \times \lsq 0, T \rsq$ into a grid with equal spacings $h = \Delta x$ and $\kappa = \Delta t$, the analytical solutions $u^{(n)} \lb x, t \rb$, $w^{(n)} \lb x, t \rb$ are approximated by the exact solution of the finite difference schemes $u^{(n)} \lb ih, j \kappa \rb$, $w^{(n)} \lb ih, j \kappa \rb$, denoted $u_{i,j}^{(n)}$ and $w_{i,j}^{(n)}$ respectively. We make use of first-order and second-order central difference approximations in the main equations for $u_{x}^{(n)}$, $u_{xx}^{(n)}$, $u_{tt}^{(n)}$ and the approximation for $u_{ttxx}^{(n)}$ can be obtained iteratively using the approximations for $u_{tt}^{(n)}$ and $u_{xx}^{(n)}$. Similar approximations are used for $w$. To simplify the obtained expressions, we introduce the notation
\begin{equation*}
D_{xx} \lb u_{i,j}^{(n)} \rb = u_{i+1,j}^{(n)} - 2u_{i,j}^{(n)} + u_{i-1,j}^{(n)}.
\end{equation*}
Using the finite-difference approximations in system \eqref{syshomog} and making use of the notation above, we obtain a tri-diagonal system of the form
{\small 
\begin{align}
&- 2 \epsilon u_{i+1,j+1}^{(2)} + \lb 4\epsilon + h^2 \rb u_{i,j+1}^{(2)} - 2 \epsilon u_{i-1,j+1}^{(2)} =  2h^2 u_{i,j}^{(2)} \notag \\
&- \frac{6 \epsilon \kappa^2}{h} \lsq \lb u_{i+1,j}^{(2)} \rb^2 - \lb u_{i-1,j}^{(2)} \rb^2 - 2u_{i+1,j}^{(2)} u_{i,j}^{(2)} + 2u_{i,j}^{(2)} u_{i-1,j}^{(2)} \rsq \notag \\
& + \lb \kappa^2 - 4 \epsilon \rb D_{xx} \lb u_{i,j}^{(2)} \rb + 2 \epsilon u_{i+1,j-1}^{(2)} - \lb 4 \epsilon + h^2 \rb u_{i,j-1}^{(2)} \notag \\
&+ 2 \epsilon u_{i-1,j-1}^{(2)} - \epsilon h^2 \kappa^2 \delta \lb u_{i,j}^{(2)} - w_{i,j}^{(2)} \rb,
\label{fd_u2}
\end{align}
}and
{\small
\begin{align}
&- 2 \epsilon \beta w_{i+1,j+1}^{(2)} + \lb 4 \epsilon \beta + h^2 \rb w_{i,j+1}^{(2)} - 2 \epsilon \beta w_{i-1,j+1}^{(2)} = 2h^2 w_{i,j}^{(2)} \notag \\
&- \frac{6 \epsilon \alpha \kappa^2}{h} \lsq \lb w_{i+1,j}^{(2)} \rb^2 - \lb w_{i-1,j}^{(2)} \rb^2 - 2 w_{i+1,j}^{(2)} w_{i,j}^{(2)} + 2 w_{i,j}^{(2)} w_{i-1,j}^{(2)} \rsq \notag \\
&+ \lb \kappa^2 c^2 -  4 \epsilon \beta \rb D_{xx} \lb w_{i,j}^{(2)} \rb + 2 \epsilon \beta w_{i+1,j-1}^{(2)} - \lb  4 \epsilon \beta + h^2 \rb w_{i,j-1}^{(2)} \notag \\
&+ 2 \epsilon \beta w_{i-1,j-1}^{(2)} + \epsilon h^2 \kappa^2 \gamma \lb u_{i,j}^{(2)} - w_{i,j}^{(2)} \rb, \label{fd_w2}
\end{align}
}for $x_a < x < x_b$ and a similar system for the third section of the bi-layer. Assuming the domain can be discretised, we denote the central point as $N = \frac{x_b - x_a}{h}$ and therefore conditions \eqref{cont_a} and \eqref{cont_b} translates directly to
\begin{align}
u_{N,j+1}^{(2)} &= u_{N,j+1}^{(3)}, &w_{N,j+1}^{(2)} &= w_{N,j+1}^{(3)}. \label{fd_cont}
\end{align}
In the continuity conditions \eqref{cont2_a} and \eqref{cont2_b}, we make use of the central difference approximations presented above, and introduce `ghost points' of the form $u_{N+1,j+1}^{(2)}$, $u_{N-1,j+1}^{(3)}$, $w_{N+1,j+1}^{(2)}$ and $w_{N-1,j+1}^{(3)}$. The continuity condition relates the ghost points to each other (omitted for brevity, but the expressions are of the same form as the ones in our earlier work\cite{KT}).

As we are considering localised initial data for strains, if we take $L$ large enough then we can enforce zero boundary conditions for the strains, i.e. $u_{x} = 0$, $w_x = 0$. We note that \eqref{fd_u2} - \eqref{fd_w2} and the associated system for  $x > x_b$ are similar to the system considered in our previous work and the same method can be applied here. 
This is summarised as follows: solve each tridiagonal system \eqref{fd_u2} - \eqref{fd_w2} (and the systems for $x > x_b$) in terms of the ghost points at the central boundary, then use the expressions for the ghost points in conditions \eqref{fd_cont} and the discretisation of \eqref{cont2_a} and \eqref{cont2_b} to obtain a solvable algebraic system for the ghost points. This is then substituted back into the tridiagonal systems to obtain the solution.

\section{Numerical formulation for coupled Ostrovsky equations}
\label{sec:PS}
In previous studies, the uncoupled Ostrovsky equation\cite{OS12} and coupled Ostrovsky equations\cite{AGK13} have been solved using finite-difference and pseudospectral\cite{F96} techniques respectively. We implement a pseudospectral method using the Runge-Kutta 4$^{\mathrm{th}}$-order method for `time' stepping in the Fourier space, and the nonlinear terms are calculated in the real domain and transformed back to the Fourier space for use in the calculation.

Firstly we transform the solution interval from $\lsq -L, L \rsq$ to $\lsq 0, 2\pi \rsq$ via the transform $\tilde{\xi} = s\xi + \pi$ and $s = \pi/L$. We denote the dependent variables by $u(\xi, X)$ and $w(\xi, X)$ and, assuming unity coefficients for simplicity, we obtain (omitting tildes)
\begin{align}
\lb u_{X} + s u u _{\xi} + s^3 u_{\xi \xi \xi} \rb_{\xi} &= \frac{1}{s} \lb u - w \rb, \notag \\
\lb w_{X} +  s w_{\xi} + s w w _{\xi} + s^3 w_{\xi \xi \xi}  \rb_{\xi} &= \frac{1}{s} \lb w - u \rb.
\label{cOstSpongeT}
\end{align}
The nonlinear term is calculated in the real domain, then transformed to the Fourier space. We can rewrite the nonlinear terms by introducing the notation
\begin{equation*}
u u_{\xi} = z_{a \xi}, \quad z_{a} = \frac{u^2}{2}, \quad w w_{\xi} = z_{b \xi}, \quad z_{b} = \frac{w^2}{2}.
\end{equation*}
The solution interval is discretised by $N$ equidistant points with spacing $\Delta \xi = 2 \pi / N$, where $N$ is a power of 2. This allows us to use the Discrete Fourier Transform
{\small 
\begin{equation*}
\hat{u} \lb k, X \rb = \frac{1}{\sqrt{N}} \sum_{j=0}^{N-1} u \lb \xi_{j}, X \rb e^{-i k \xi_{j}},  \  -\frac{N}{2} \leq k \leq \frac{N}{2} - 1,
\end{equation*}
}and a similar transform for $w$. Here $k$ is an integer which represents the discretised (and scaled) wavenumber. In what follows we first exclude the zero mode, i.e. we let $k \ne 0$, assuming zero mass initial conditions for $u$ and $w$. 

The inverse transform is
{\small 
\begin{equation*}
u \lb \xi, X \rb = \frac{1}{\sqrt{N}} \sum_{k=-N/2}^{N/2 - 1} \hat{u} \lb k, X \rb e^{i k \xi_{j}}, \quad j = 0, 1, \dots , N - 1,
\end{equation*}
}and again a similar transform for $w$. We make use of the Fast Fourier Transform (FFT) algorithm to implement these transforms efficiently. The discrete Fourier transform of equations \eqref{cOstSpongeT} with respect to $\xi$ gives
\begin{align}
\hat{u}_{X} + i k s \hat{z}_{a} - i k^3 s^3 \hat{u} &= -\frac{i}{k s} \lb \hat{u} - \hat{w} \rb, \notag \\
\hat{w}_{X} +  i k s \hat{w} + i k s \hat{z}_{b} - i k^3 s^3 \hat{w}  &= - \frac{i}{k s} \lb \hat{w} - \hat{u} \rb.
\label{PSEq}
\end{align}
We now introduce the Runge-Kutta 4$\mathrm{th}$-order method. Assume that the solution at $X$ is given by $\hat{u}_j = u(k, j \varkappa)$ and $\hat{w}_j = u(k, j \varkappa)$, where $\varkappa = \Delta X$. Then the solution at $X = (j+1) \Delta X$ is given by
\begin{align}
\hat{u}_{k, \lb j+1 \rb \varkappa} &= \hat{u}_{k, j \varkappa} + \frac{1}{6} \lb a_{1} + 2 b_{1} + 2 c_{1} + d_{1} \rb, \notag \\
\hat{w}_{k, \lb j+1 \rb \varkappa} &= \hat{w}_{k, j \varkappa} + \frac{1}{6} \lb a_{2} + 2 b_{2} + 2 c_{2} + d_{2} \rb,
\label{uwSolRK4}
\end{align}
where $a_i, b_i, c_i, d_i$ are functions of $\xi$ at a given moment in time, $X$, and are defined as
{\small 
\begin{align*}
&a_{i} = \varkappa F_{i} (\hat{u}_{j}, \hat{w}_{j}), &&b_{i} = \varkappa F_{i} (\hat{u}_{j} + \frac{a_{1}}{2}, \hat{w}_{j} + \frac{a_{2}}{2}), \\
&c_{i} = \varkappa F_{i} (\hat{u}_{j} + \frac{b_{1}}{2}, \hat{w}_{j} + \frac{b_{2}}{2}), &&d_{i} = \varkappa F_{i} 1(\hat{u}_{j} + c_{1}, \hat{w}_{j} + c_{2}),
\end{align*}
}for $i=1,2$. The functions $F_{i}$ are found as a rearrangement of \eqref{PSEq} to contain all non-time derivatives. Explicitly we have
\begin{align*}
F_{1} \lb \hat{u}_{j}, \hat{w}_{j} \rb =& - i k s \hat{z}_{a} + i k^3 s^3 \hat{u_j} - \frac{i}{k s} \lb \hat{u_j} - \hat{w_j} \rb, \notag \\
F_{2} \lb \hat{u}_{j}, \hat{w}_{j} \rb =& - i k s \lb \hat{z}_{b} + \hat{w_j} \rb + i k^3 s^3 \hat{w_j} - \frac{i}{k s} \lb \hat{w_j} - \hat{u_j} \rb.
\end{align*}

For the cases in Section \ref{sec:FinDelam}, where the waves re-enter a coupled region after a delamination, we need to introduce a linear damping region (`sponge layer') and add this at each end of the domain to prevent radiated waves re-entering the region of interest and interfering with the main wave structure.\cite{AGK13} The sponge layer is defined as 
{\small
\begin{equation}
r(x) = \frac{\nu}{2} \lsq \tanh{K \lb x - \frac{3L}{4} \rb} - \tanh{K \lb x + \frac{3L}{4} \rb} \rsq,
\label{Sponge}
\end{equation}
}for some constants $\nu, K$. We choose $K$ so that $K L = 12$ and $\nu$ is chosen so that damping occurs quickly. The sponge layer is incorporated as
\begin{align}
\lb u_{X} + u u _{x} + u_{x x x} + r(x) u \rb_{x} &= \delta \lb u - w \rb, \notag \\
\lb w_{X} + w_x + w w _{x} + w_{x x x} + r(x) w \rb_{x} &= \gamma \lb w - u \rb,
\label{cOstSponge}
\end{align}
and 
is treated in the same way as nonlinear terms.

To remove aliasing effects, we use the truncation 2/3-rule by Orszag in Boyd.\cite{Boyd01} This effect is due to the pollution of the numerically calculated Fourier transform by higher frequencies due to the series being truncated. 

Finally, for the particular non-zero mass initial conditions used in this paper, in the Fourier space the zero mode (for $k = 0$) is approximated as a small constant  $\O {10^{-2}}$, defined by the initial condition. The maximum of the zero mode of $u-w$ is $\O {10^{-3}}$  for the cases considered in the paper, and therefore this approximation introduces only a small error, approximately satisfying the equations (\ref{PSEq}) (multiplied by $k$) for zero modes. In the real space, the average $\frac{1}{2L} \int_{-L}^{L} (u-w) d \xi$ is $\O{10^{-5}}$, showing that the zero mass constraint is approximately satisfied. A more accurate approach is required for general initial conditions with non-zero mass.\cite{KMP14}


\end{document}